\newcommand{\pd}[2]{ { \partial {#1} \over \partial {#2} } }
\newcommand{\pdd}[2]{ { \partial^2 {#1} \over \partial {#2}^2 } }
\newcommand{\pddd}[3]{ { \partial^2 {#1} \over \partial {#2} \partial {#3} } }

\def\half{\frac{1}{2}}
\def\ptl{\partial}
\def\beqar{\begin{eqnarray}}
\def\eeqar{\end{eqnarray}}
\def\ta{\tilde{A}}
\def\td{\tilde{D}}
\def\tg{\tilde{\gamma}}
\def\g{\gamma}
\def\a{\alpha}
\def\b{\beta}
\def\l{\lambda}

\newcommand{\llabel}[1]{\label{#1}}              

\newcommand{\labeq}[2]{ \begin{equation} \llabel{#1}
{#2}
\end{equation}}

\documentclass[aps,showpacs,twocolumn,prd,preprintnumbers,amsmath,amssymb,letterpaper]{revtex4}

\usepackage{amsmath}

\begin{document}

\title{\bf Well-posed constrained evolution of 3+1 formulations of General Relativity}

\author{Vasileios Paschalidis${}^1$, Alexei Khokhlov${}^{1,2}$, and Igor Novikov${}^{2,3,4}$}

\affiliation{${}^1$ Department of Astronomy and Astrophysics, The
University of Chicago, 5640 S Ellis Ave., Chicago IL 60637  \\
${}^2$ Enrico Fermi Institute, The University of Chicago, 5640 S
Ellis Ave., Chicago, IL 60637 \\
${}^3$ Niels Bohr Institute, Blegdamsvej 17, DK-2100
Copenhagen, Denmark \\
${}^4$ Astro Space Center of P.N. Lebedev Physical Institute,
Profsoyuznaya 84/32, Moscow, 117810, Russia}

\date{\today}

\begin{abstract}
We present an analysis of well-posedness of  constrained evolution
of 3+1 formulations of GR. In this analysis we explicitly take
into account the energy and momentum constraints as well as
possible algebraic constraints on the evolution of high-frequency
perturbations of solutions of Einstein's equations. In this
respect, our approach is principally different from standard
analyses of well-posedness of free evolution in general
relativity. Our study reveals the existence of subsets of the
linearized Einstein's equations that control the well-posedness of
constrained evolution. It is demonstrated that the well-posedness
of ADM, BSSN and other 3+1 formulations derived from ADM by adding
combinations of constraints to the right-hand-side of ADM and/or
by linear transformation of the dynamical ADM variables depends
entirely on the properties of the gauge. For certain classes of
gauges we formulate conditions for well-posedness of constrained
evolution. This provides a new basis for constructing stable
numerical integration schemes for a classical
Arnowitt--Deser--Misner (ADM) and many other 3+1 formulations of
general relativity.

\end{abstract}



\maketitle



\section{Introduction}

An outstanding problem of numerical general relativity (GR) is
achieving long-term stable numerical integration of Einstein's
equations. Until very recently, problems such as the general case
of colliding black holes (BH) could not be solved due to
instability of numerical integration \cite{2,3,4,5,6,7}. During
the last year, several groups
 have succeeded in simulating certain binary black hole (BBH) spacetimes.
Collisions of non-rotating black holes were simulated
 using BSSN \cite{3} with a specific choice of ``$1+\log$'' slicing and
modified gamma-freezing shift conditions in \cite{1a} and
\cite{1b}. A three-dimensional  collision of two black holes
 originated from the collapse of a scalar field was simulated using a generalized
harmonic decomposition of the GR field equations in \cite{1c}. In
addition to using a special gauge, the authors of \cite{1a} and
\cite{1b} enforced  some of the BSSN constraints and used some of
the  constraints to control the constraint violating modes. The
author of \cite{1c}, in addition to special coordinates he used
the constraints to damp constraint violating modes.

In spite of this remarkable success the general problem of
long-term and stable integration of the Einstein equations remains
unsolved. There is no general method to choose an appropriate
formulation and appropriate coordinates to guarantee
well-posedness and stability of numerical integration for a
general strong field case. In particular, one does not know how
the approaches used in \cite{1a,1b,1c} will behave in other strong
vacuum field astrophysical cases, or cases where matter or even
matter and magnetic fields are present. Perhaps the final word to
this problem would be the development of schemes which implement
fully constrained evolution of well-posed formulations. The
purpose of this paper is to formulate a general approach to study
the well-posedness of constrained evolution of 3+1 formulations of
GR.

Generally, a 3+1 formulation is comprised of the evolutionary part
and the constraints. The standard approach to solving a 3+1 system
consists of integration of the evolutionary part in time (free
evolution) starting with constrained initial conditions. If the
constraints are satisfied initially, they should automatically be
satisfied throughout the evolution due to the mathematical
properties of Einstein's equations.

Well-posedness of free evolution of 3+1 formulations has been
analyzed in \cite{4,13}. For example, a classical ADM 3+1
formulation \cite{2} is usually ill-posed. Ill-posedness of free
evolution precludes stable numerical integration. We note,
however, that well-posedness can not guarantee global in time
existence of solutions, but only local existence. However, it is a
necessary condition for stable integration.

There has been a number of attempts to overcome the instability of
numerical integration. Using a harmonic gauge makes the ADM 3+1
system well-posed \cite{8}. However, this gauge is not convenient
for many physical problems. Introduction of a conformal factor and
the trace of extrinsic curvature as additional unknown variables
into the system allows to increase the duration of stable
integration \cite{3}. The evolutionary part of a 3+1 system  can
also be modified by adding a combination of constraints to its
right-hand side. Choosing a special gauge (densitized lapse and
zero shift) and addition of certain combinations of constraints to
the right hand sides (RHS) of the ADM evolution equations makes
the modified system well-posed \cite{4}. However, all these
modifications did not lead to a complete elimination of
instabilities. Numerical experiments show that in general
three-dimensional problems of GR the constraint equations are
eventually violated even for a well-posed free evolution, and this
terminates computations. Little progress has been made towards a
theoretical understanding of this behavior. Possible explanations
are given in \cite{7,13}.

Recently, attempts have been made to improve the behavior of
modified 3+1 systems by enforcing the constraints after every
integration time step of a free evolution \cite{5,6}. According to
\cite{5} this procedure allows integration of an isolated
spherical black hole space-time for extended periods of time. We
also mention that for certain cases perturbative approaches
provide an alternative to straightforward numerical integration,
e.g., for forming initial conditions for BH collisions \cite{10}.
Yet, the general problem of long-term stable integration remains
open.

In a high-frequency perturbation analysis of a free evolution it
is possible to separate perturbations on three parts: (1)
perturbations of space-time itself, (2) perturbations of a
coordinate system, and (3) perturbations describing deviations
from constraints. If the behavior of space-time at a given point
does not depend on future, then we must associate ill-posedness
with coordinate and constraint-violating modes of perturbations.
To achieve stable numerical integration, we must (A) use a gauge
that does not lead to ill-posedness, and (B) eliminate or suppress
ill-posedness caused by constraint violating modes.

A general theory of gauge stability (problem A) has been
formulated, and well-posedness of gauges has been analyzed in
\cite{11}. It was demonstrated that coordinate perturbations in
the evolution of the metric can be separated from the other two
types of perturbations and the study of gauge stability can be
reduced to a study of a general quasi-linear system of eight
coupled partial differential equations for perturbations of lapse
$\alpha$, shift $\beta_i$, $i=1,...,3$, and perturbations of
space-time coordinates $x^a$, $a=0,...,3$. Conditions for
well-posedness have been formulated in \cite{11} for several types
of gauges. We will repeatedly return to this subject in subsequent
sections.

Constraint-violating modes of perturbations (problem B) are not
fully understood at present. Recently, attempts have been made to
stabilize numerical integration by enforcing the constraint
equations after every integration time step of a hyperbolic free
evolution \cite{5,6}. Such enforcement is not a unique procedure.
Several possibilities are discussed in \cite{6}. According to
\cite{5}, constraint enforcement improves integration of an
isolated spherical black hole space-time. Analysis of
well-posedness of constraint enforcing procedure for a hyperbolic
3+1 formulation, densitized lapse, zero shift, and flat Minkowski
space-time is given in \cite{9}.

An alternative to enforcement of constraints after a free
evolution time step may be the construction of numerical schemes
for constrained evolution in which growing constraint-violating
modes are explicitly removed. In order to achieve this goal we
must understand the nature of evolution of perturbations which
satisfy constraints.

In this paper we present a general analysis of
constraint-satisfying perturbations and address the issue of
well-posedness of constrained evolution of 3+1 formulations of GR.
We explicitly take into consideration the energy and momentum
constraints on the evolution of high-frequency perturbations of
solutions of Einstein's equations. In this respect, our analysis
is principally different from standard analyses of well-posedness
of a free evolution in general relativity. Our study reveals the
existence of subsets of the linearized version of Einstein's
equations that control well-posedness of constrained evolution. We
demonstrate that the well-posedness of ADM, BSSN and other 3+1
formulations derived from ADM by adding combinations of
constraints to the right-hand-side (RHS) of ADM and/or by linear
transformation of the dynamical ADM variables depends entirely on
the properties of the gauge. For certain classes of gauges we
formulate conditions of well-posedness. The existence of these
subsets provides a basis for constructing stable numerical
integration schemes that incorporate the constraints directly.

The paper is organized as follows. We begin with the ADM 3+1
formulation and gauge classification (Section II). In Section III
we give a general theory of well-posedness of constrained
evolution of ADM. In Section IV we extend our theory to other 3+1
formulations including the Kidder-Scheel-Teukolsky (KST) and the
Baumgarte-Shapiro-Shibata-Nakamura (BSSN). Discussion and
conclusions are given in Section V.

\section{ADM 3+1 formulation and Gauges}

A general form of the ADM 3+1 formulation consists of the
evolutionary part
\labeq{ADM-gamma}{ \pd{\gamma_{ij}}{t} = -2\alpha K_{ij} +
\nabla_i\beta_j + \nabla_j \beta_i, }
\labeq{ADM-K}{
\begin{split}
\pd{K_{ij}}{t}  = & \alpha \left( ^{(3)}R_{ij} +  K K_{ij} - 2
\gamma^{mn} K_{im} K_{jn} \right)
               -\nabla_i \nabla_j \alpha \\
             &  + (\nabla_i\beta^m) K_{mj} + (\nabla_j\beta^m) K_{mi} + \beta^m \nabla_m K_{ij},\\
\end{split}
}
and the energy and momentum constraints which we will call the
kinematic constraints,
\labeq{ADM-H}{
               {\cal H}:\quad  ^{(3)}R + K^2 - K_{mn} K^{mn} = 0,
}
\labeq{ADM-M}{
              {\cal M}:\quad  \nabla_m K^{m}{}_i - \nabla_i K = 0, \quad i=1,2,3,
}
where  $K = \gamma^{mn} K_{mn}$, $\gamma_{ij}$ and $K_{ij}$ are
the three-dimensional metric of a space-like hypersurface and the
extrinsic curvature respectively, $\alpha $ is the lapse function,
$\beta^i$ is the shift vector (these are gauge functions), and
$^{(3)}R_{ij}$ is the three-dimensional Ricci tensor, $^{(3)}R =
\gamma^{ij} {^{(3)}R_{ij}}$. We must add a specification of gauge
(lapse and shift) in order to close the system \eqref{ADM-gamma},
\eqref{ADM-K}.

In this paper we use a general gauge specification similar to that
introduced in \cite{11} with one modification. Instead of working
with the dual shift vector $\beta_k$ here we work with the shift
vector $\beta^k = \gamma^{kj} \beta_j$. A general gauge can be
specified as
\labeq{Gauge}{\begin{split} F_a\bigg (      x^b, \alpha, \beta^i,&
\pd{\alpha}{x^b},\pddd{\alpha}{x^b}{x^c}, ..., \pd{\beta^i}{x^b}
,...
      \gamma_{ij}, \pd{\gamma_{ij}}{x^b},...
\bigg) =0, \\ & a,b,c = 0,...,3, \quad i,j = 1,2,3, \end{split}}
where it is implied by the ellipsis that one can add
higher order derivatives of both the lapse and the shift
vector, e.g., $\pddd{\beta^i}{x^a}{x^b}$ or the dynamical
variables, e.g., $\pddd{K_{ij}}{x^a}{x^b}$, $\pddd{\gamma_{ij}}{x^a}{x^b}$ , and so on.
Following \cite{11}, we distinguish three types of gauges.

\medskip\noindent
1. {\it Fixed gauges} for which both the lapse and shift are
functions of coordinates $t=x^0$ and $x^i, i=1,...,3$ only,
\labeq{FixedGauge}{
              \alpha=\alpha(t, x^i), \quad
              \beta^k=\beta^k(t,x^i), \quad i,k=1,2,3.
}
Geodesic slicing $\alpha=1$, $\beta^k=0$ is a specific case of a
fixed gauge.

\medskip\noindent
2. {\it Algebraic (local) gauges} for which both the lapse and
shift can be expressed as algebraic functions of coordinates and
local values of $\gamma_{ij}$ and its derivatives,
\labeq{AlgebraicGauge}{
                          \alpha = \alpha \left( x^a, \gamma_{ij}, \pd{\gamma_{ij}}{x^b}, ... \right), \quad
                          \beta^k = \beta^k \left( x^a, \gamma_{ij}, \pd{\gamma_{ij}}{x^b}, ... \right).
}

\medskip\noindent
3. {\it Differential (non-local) gauges} which are defined by a
set of partial differential equations and which cannot be reduced
to an algebraic form. Algebraic gauges are a subset of
differential gauges. We call the differential gauges non-local,
because the gauge variables are not completely defined by the
local values of the dynamical variables. Differential gauges are
governed by differential equations and hence the lapse and shift
may also depend on boundary conditions.

For fixed and algebraic gauges, the total number of partial
differential equations of the ADM formulation does not increase
compared to \eqref{ADM-gamma}- \eqref{ADM-M}. For differential
gauges, a complete ADM formulation will consist of
\eqref{ADM-gamma} - \eqref{ADM-M} plus differential equations
describing the gauge.

\section{Analysis of well-posedness}

We begin with a brief description of our general approach to the
analysis of well-posedness of constrained sets of partial
differential equations (PDEs). Let
\labeq{def}{
\partial_t \vec{u}
= {\hat M}^{\ell}(\vec{u}) \, \partial_{\ell} \vec{u} +{\vec
M}^o(\vec{u}), \quad \ell=1,2,3}
be a set of n first order quasi linear partial differential
equations, where $\vec{u}$ is the column vector of the $n$ unknown
variables, ${\hat M}^\ell$ are $n\times n$ matrices and ${\vec
M}^o$ is a $n$-component column vector.

The concept of mathematical well-posedness is often related to
strong hyperbolicity. For system \eqref{def} we present the
following theorem from \cite{21} without proof.

\paragraph{\textbf{Theorem 1.}}

The Cauchy problem for a first-order system of quasi-linear PDEs
\eqref{def} is well posed if and only if the following two
conditions hold:

\begin{enumerate}
\item For all unit one forms $v_i$, all eigenvalues of the
characteristic matrix, $\hat M=v_i \hat M^i$ are purely real.

\item There is a constant $K$, and for each $v_i$, there is a
transformation $\hat S(v_i)$ with
\labeq{detcon2}{|\hat S(v_i)|+|\hat S^{-1}(v_i)|\leq K,}
such that the transformed matrix $\hat S(v_i) \hat M(v_i) \hat
S^{-1}(v_i)$ is diagonal.
\end{enumerate}

\paragraph{\textbf{Definition 1.}} A first-order system of
quasi-linear PDEs \eqref{def} is called strongly hyperbolic if all
conditions of the theorem above are met.

\paragraph{\textbf{Definition 2.}} A first-order system of
quasi-linear PDEs \eqref{def} is called weakly hyperbolic if it
satisfies only the first condition of Theorem~1 and does not
satisfy the second condition. Weakly hyperbolic systems are
ill-posed.

For a constrained system \eqref{def}, the dynamical variables
satisfy a set of $m < n$ quasi linear constraint equations of the
form: \labeq{con}{ {\hat C}^{\ell}(\vec{u}) \,
\partial_{\ell} \vec{u} +{\vec C}^o(\vec{u})=0, \quad \ell=1,2,3}
where  ${\hat C}^\ell$ are $m\times n$ matrices and ${\vec C}^o$
is an $m$-component column vector . We call a constrained surface
a collection of all solutions of \eqref{def} which satisfy the
constraints. If evolution starts on the constrained surface it
will remain on this surface.

In order to study well-posedness one drops the zeroth-order terms $\vec M^o$, freezes
the coefficients $\hat A^i$ of \eqref{def} and studies the characteristic matrix
of the system for all frozen in problems. This is equivalent to considering
 1) high frequency and 2) small amplitude planar perturbations on the
dynamical variables along a line locally specified by a unit
vector $\vec{v}$ and parameterized by $\lambda$,  so that for an
arbitrary function $u(x^1,x^2,x^3)$
\labeq{direction}{ \frac{\partial{u}}{\partial
x^i}=v_i\frac{\partial{u}}{\partial\lambda}, } where $v_i$ is the
dual vector to $v^i$, given by $v_i=\g_{ij}v^j$ and $\g_{ij}$ is
the 3-metric of a spacelike hypersurface embedded in the manifold
carrying the background solution $\vec{u}$ about which we perturb.
The two aforementioned properties are the properties of all
perturbations considered in this paper and those will be implied
whenever the word ``perturbation" is used.
For perturbations $\delta\vec{u}$ on $\vec{u}$, combination of \eqref{def} and \eqref{direction} yields
\labeq{deflong1}{
\partial_t \delta \vec{u}
= {\hat M}^{\ell}v_{\ell} \pd{\delta \vec{u}}{\lambda}\equiv{\hat
M}\pd{\delta \vec{u}}{\lambda}. }
For perturbations of $\vec{u}$ to remain on the constraint surface,
they must satisfy the linearized constraint equations. After
linearizing equations \eqref{con} we obtain
\labeq{conlong1}{ {\hat C}^{\ell}v_\ell \, \pd{\delta
\vec{u}}{\lambda} \equiv\hat{C} \pd{\delta \vec{u}}{\lambda}=0, }
where ${\hat M}$ and ${\hat C}$ are the principal matrices of
equations \eqref{def} and \eqref{con} respectively. Equations
\eqref{conlong1} is a set of $m$ equations for the spatial
derivatives of the $n$ unknown variables, which in general can be
solved for $m$ of the $n$ spatial derivatives of variables
$\vec{u}$. Substitution of \eqref{conlong1} into \eqref{deflong1},
leads to a set of $q=n-m$ linear partial differential equations
for $q$ of the initial $n$ variables. This
 is schematically given by
\labeq{minimal}{
\pd{\vec{a}_{q}}{t}=\hat{A}_q(\vec{u},v_i)\pd{\vec{a}_q}{\lambda},
}
where $\hat{A}_q$ is a $(q\times q$) matrix. We will refer to
\eqref{minimal} as the minimal set. The solution of
\eqref{minimal} completely  determines the solution of the entire
linearized system \eqref{deflong1}. Therefore, the well-posedness
of the minimal set determines the well-posedness of the entire
constrained system. Theorem~1 and Definitions~1,2 apply to
\eqref{minimal}.

We should note here that the description above is for first-order
systems of PDEs. When one deals with second-order systems, then it
is straightforward to obtain the equivalent first-order system of
PDEs, by simply defining the first-order derivatives as new
variables. Courant and Hilbert \cite{19} show that if one derives
the first-order form in the fashion described above, then the
totality of solutions of the two systems coincide, for given
Cauchy data. In addition to that in \cite{20} it is shown that the
hyperbolic properties of the second-order system and its
equivalent first-order counterpart are the same. Although a
reduction to a first-order system is not necessary, it makes it
easier to analyze well-posedness, since in second-order systems
one has to carefully study the behavior of the first order terms.

\subsection{Linearized equations of ADM}

We want to study well-posedness of an ADM 3+1 formulation
(evolutionary part \eqref{ADM-gamma},\eqref{ADM-K} plus
constraints \eqref{ADM-H}, \eqref{ADM-M} ).
%
For the analysis of well-posedness it is convenient to rewrite
equations \eqref{ADM-gamma} - \eqref{ADM-M} in first-order form.
We introduce new variables
\labeq{D-defin}{
                      D_{ij;k} \equiv \pd{\gamma_{ij}}{x^k}
}
and drop all low-order terms, which do not contribute to the
principal part of the equations. In terms of these variables, the
ADM equations linearized with respect to a certain unperturbed
solution (not necessarily a flat Minkowski spacetime)
\labeq{Base}{
                \gamma_{ij}, \quad K_{ij}, \quad D_{ij,k}, \quad \alpha, \quad \beta^k
}
can be written as
\labeq{ADM-gamma-D}{ \pd{\delta\gamma_{ij}}{t}  =  2
\gamma_{\ell(i}\ptl_{j)}\delta\beta^\ell, }
\labeq{dKij}{\begin{split} \pd{\delta K_{ij}}{t}  =
\alpha\bigg(&-\frac{1}{\alpha}\ptl_i\ptl_j\delta\alpha+
R_{ij}^{1}+
                   \frac{\beta^k}{\alpha} \pd{\delta K_{ij}}{x^k}
                   \bigg) \\
                   & +\Gamma^k_{ij}\ptl_k\delta\alpha+2
                   K_{k(i}\ptl_{j)}\delta\b^k, \end{split}
}
\labeq{dDijk}{ \begin{split} \pd{\delta D_{ij;k}}{t}  = \alpha
\bigg(&-2\pd{\delta
K_{ij}}{x^k}+\frac{\beta^{\ell}}{\alpha}\pd{\delta
D_{ij;k}}{x^\ell} \bigg)
                -2K_{ij}\pd{\delta\alpha}{x^k}
                \\ & +2\ptl_k[\gamma_{\ell(i}\ptl_{j)}\delta \beta^\ell]
                +D_{ij;\ell}\pd{\delta\beta^\ell}{x^k},
\end{split}
}
where $\delta \gamma_{ij}$, $\delta K_{ij}$, $\delta D_{ij;k}$,
$\delta \alpha$, and $\delta\beta^m$ are perturbations of
\eqref{Base},  $R_{ij}^{1}$ is the principal part of the Ricci
tensor
\labeq{R1}{\begin{split}
   R^1_{ij} = &\frac{1}{2} \gamma^{sk} \left( \pd{\delta D_{ik;j}}{x^s}
   + \pd{\delta D_{jk;i}}{x^s} - \pd{\delta D_{ij;k}}{x^s}  \right)
             \\ &   - \frac{1}{2} \gamma^{mn} \pd{\delta
             D_{mn;i}}{x^j},\end{split}
}
and
\labeq{Gamma}{
                  \Gamma^k_{ij} = \frac{1}{2} \gamma^{kn} \left( D_{in;j} + D_{jn;i} - D_{ij;n}  \right).
}
To close the system equations \eqref{ADM-gamma-D} - \eqref{MD}
must be supplemented with the linearized version of gauge
equations \eqref{Gauge}.

We notice that the evolution equations for the 3-metric \eqref{ADM-gamma-D} are decoupled from the
evolution equations for $K_{ij}$, $D_{ij;k}$ and the linearized gauge equations. It is the subsystem \eqref{dKij}, \eqref{dDijk} and the
linearized gauge equations which determine the well-posedness of the entire system. The
solution of \eqref{ADM-gamma-D} is completely determined by the solution of the above subsystem.

The linearized constraint equations in new variables are
\labeq{HD}{
          {\cal H}:\quad   R^1 = \gamma^{nm}R^1_{nm} = 0,
}
and
\labeq{MD}{
          {\cal M}_i:  \quad   \gamma^{ms}  \pd{\delta K_{mi}}{x^s} -
          \delta^s_i \gamma^{mn} \pd{\delta K_{mn}}{x^s}  = 0 ,
}
where $\delta^s_i$  is the Kronecker symbol. The introduction of
extra variables imposes new linear constraints on the system,
which, for perturbations of \eqref{Base}, can be written as
follows:
\labeq{DConstraints}{
                                     \pd{\delta D_{ij;k}}{x^m} = \pd{\delta
                                     D_{ij;m}}{x^k},
}
where \eqref{DConstraints} is derived by use of \eqref{D-defin}
and the fact that partial derivatives commute.

\subsection{Analysis of well-posedness of constrained evolution and well-posed subsets}

We consider planar perturbations
\labeq {Planar}{ \delta \gamma_{ij},\quad  \delta K_{ij},\quad
\delta D_{ij;k}, \quad \delta \alpha,\quad \delta\beta^m, }
moving in an arbitrary direction locally specified by a unit
vector $v^i$, $v^i v_i = 1$. Substitution of \eqref{Planar} into
\eqref{ADM-gamma-D} - \eqref{dDijk} gives a set of thirty linear
PDEs for perturbations \eqref{Planar}. Substitution of
\eqref{Planar} into the eighteen in number PDEs
\eqref{DConstraints} gives twelve independent linear PDEs for
those perturbations. This means that along the direction which we
are probing only six components of the eighteen $\delta D_{ij;k}$
are independent. The linearized energy and momentum constraints
\eqref{HD} and \eqref{MD} give four additional linear PDEs for
perturbations \eqref{Planar}. None of these equations contain time
derivatives of \eqref{Planar}, because the constraints do not
involve time derivatives of the perturbations.

Here we note that in the context of the 1st-order formulation
equations \eqref{D-defin} are constraints on the initial
data. They are not involved in the analysis of well-posedness
but they serve to guarantee the coincidence of the solutions of the 1st order and second order ADM equations.
Since the evolution of the perturbations of the 3-metric is decoupled from and determined by the evolution of the perturbations
of the subset variables $K_{ij}$, $D_{ij;k}$ and the linearized gauge equations,
the study of well-posedness of a constrained evolution reduces to analyzing the subset of
variables $K_{ij}$, $D_{ij;k}$, and all remaining
constraints \eqref{HD}, \eqref{MD}, \eqref{DConstraints}.
There are twenty four evolution equations and sixteen constraints involved that
leads  to eight degrees of freedom.

By elimination of 16 spatial derivatives of
the dynamical variables from the subset of twenty four equations, we obtain
eight linearly independent equations, which form the
minimal set. This can be schematically written as
\labeq{A8}{
             \pd{ \vec a_8}{t} = \hat A_8(\vec{u},v_i) \pd{\vec
             a_8}{\lambda},
} where  $\hat A_8$ is an $8\times 8$ matrix which depends on the
direction of planar perturbations and the background solution,
$a_8$ are the independent perturbation amplitudes.

To study the well-posedness of the Cauchy problem of  system
\eqref{A8} we first find the eigenvalues $\omega_k$ of the
principal matrix of the system for every direction $v_i$. If
non-zero imaginary parts are present in some of those eigenvalues,
this indicates that part of the system forms an elliptic subset
and the Cauchy problem is ill-posed. If all $\omega_k$ are real,
then we investigate whether $\hat A_8$ is diagonalizable for every
direction $v_k$ (uniformly diagonalizable), and whether the second
condition \eqref{detcon2} of Theorem~1 is satisfied
(transformation $\hat S$ is uniformly bounded).

A convenient way to investigate properties of $\hat A_8$ is to use
a reduction of $\hat A_8$ to a Jordan canonical form \cite{15},
\labeq{JORDAN}{
       \hat A_8 = \hat S \hat J \hat S^{-1},
}
so that \labeq{A8a}{ \quad \pd{}{t} \left( \hat S^{-1}  \vec a_8
\right) = \hat J \pd{}{\lambda}\left( \hat S^{-1} \vec a_8
\right), } where $\hat S$ is a non-singular matrix of a similarity
transformation \eqref{JORDAN} and $\hat J$ is a block-diagonal
Jordan canonical matrix
\labeq{Jordan}{
                  \hat J = \begin{bmatrix}
                                               \hat J_1 & 0 & ... & 0 \\
                                               0        &\hat J_2 & ... & 0\\
                                               ...      &  ...    & ... & ... \\
                                               0      &  0    & ... & \hat J_N \\
\end{bmatrix}.
}
consisting of canonical Jordan blocks $\hat J_k$. Each canonical
block $\hat J_k$ has the form

\labeq{JBlock}{
           \hat   J_k  =   \begin{bmatrix} \omega_k & 0 & 0 & ... & 0 & 0\\
                                    1       & \omega_k  & 0 & ... & 0 & 0\\
                                   0        & 1 & \omega_k   & ... & 0 & 0\\
                                   ...      & ... & ... & ... & ... & ...\\
                                   ...      & ... & ... & ... & 1 & \omega_k\\
\end{bmatrix}
}
In general, the number of canonical blocks $N$ may vary from one
to eight and the size of each square block can vary from one to
eight as well. The diagonal elements of $\hat J$ are eigenvalues
of $\hat A_8$. If for any and all possible directions $v_i$ all
Jordan blocks have size one (i.e. all off-diagonal elements of
$\hat J$ are zero), then $\hat A_8$ has a complete set of
eigenvectors and hence $\hat A_8$ is uniformly diagonalizable.

If some of off-diagonal elements of $\hat J$ are non-zero at least
in one direction $v_k$, the set of eigenvectors in this direction
is not complete and hence \eqref{A8} cannot admit a well-posed
Cauchy problem. The system \eqref{A8} is then weakly hyperbolic.

In general, uniform diagonalizability of $\hat A_8$ may not be
enough to guarantee the existence of a uniformly bounded
transformation $\hat S$ required by Theorem~1. We show in Appendix
A, however, that if the eigenvalues $\omega_k$ of $\hat A_8$ are
analytic functions and the elements of $\hat A_8$ are ratios of
analytic functions, then uniform diagonalizability guarantees the
existence of a uniformly bounded transformation matrix $\hat S$.
For the general relativity cases studied in the paper, the
elements of the matrices of all minimal sets are always ratios of
polynomials in $v_k$. Furthermore, those matrices have real
eigenvalues which are analytic functions of $v_k$. Thus, for those
systems studied here uniform diagonalizability is sufficient in
order to satisfy the conditions for well-posedness.

The Jordan decomposition also provides information about
combinations  of independent variables that evolve according to
the corresponding eigenfrequencies $\omega_k$. These combinations
can be determined by applying a similarity transformation $\hat
U^{-1}$ to the vector of original variables $\vec a_8$
\eqref{A8a}. In the case of week hyperbolicity or elliptic
behavior, this allows one to find a combination of variables whose
time evolution is responsible for this behavior.

The above discussion is strictly valid for ADM and any 3+1
formulation of general relativity derived from ADM by a linear
transformation of variables and addition of combinations of
constraints on the RHS of the ADM equations, provided that they
are coupled with fixed or algebraic gauges. If a gauge is
differential (specified by a general \eqref{Gauge}), then the
number of linearized ADM equations may be greater than thirty and
the final system of independent perturbations \eqref{A8} will
contain more than eight components. Analogous consideration must
then be applied to this larger reduced system. A detailed
discussion of well-posedness of ADM 3+1 coupled with differential
gauges is given in section~\ref{ADM_plus_Dif_Gauges}. Next section
discusses well-posedness of ADM with fixed and algebraic gauges.

\subsection{Well-posed subsets for ADM with fixed and algebraic gauges}

For fixed and algebraic gauges, the study of \eqref{A8}  can be
carried out analytically but resulting expressions for an
arbitrary  direction $v_k$ are extremely complicated. We present
here only results  for algebraic gauges of a simple form
\labeq{gauge1}{
           \alpha = \alpha(t, x^k, \gamma_{ik}), \quad \beta^i = \beta^i(t, x^k)
}
and
\labeq{gauge2}{
           \alpha = \alpha(t, x^k, \gamma_{ik}), \quad \beta_i = \beta_i(t, x^k).
}
For gauge \eqref{gauge1} the perturbation frequencies obtained
from \eqref{A8}  are

\labeq{gauge1-freq}{\begin{split}
            \omega_{1,2} =&\b^iv_i, \quad
            \omega_{3,...,6} = \b^iv_i\pm \alpha, \\
            \quad \omega_{7,8} = &\b^iv_i \pm  \sqrt{  \pd{\alpha^2}{\gamma_{ij}} v_i
            v_j}\end{split}
}
and for gauge \eqref{gauge2} they are
\labeq{gauge2-freq}{\begin{split}
            \omega_{1,...,6} = &\pm \sqrt{ \alpha^2 - \beta_i\beta^i
            }, \\
             \quad \omega_{7,8} = &\frac{\beta^i v_i \pm
                                                      \sqrt{\left( \pd{\alpha^2}{\gamma_{ij}}
                                                      + \beta^i\beta^j\right) v_i v_j}
                                                      }{2}.
\end{split}}
The difference in formulas \eqref{gauge1-freq} and
\eqref{gauge2-freq} arises because \eqref{gauge1} fixes the shift
vector whereas \eqref{gauge2} fixes its dual counterpart.
Metric-independent contravariant shift corresponds to
metric-dependent covariant shift and vice-versa. The results for
gauge \eqref{gauge2} coincide exactly with those of our analysis
of this gauge in \cite{11}.

Six eigen-frequencies $\omega_{1,...,6}$ in  \eqref{gauge1-freq}
and \eqref{gauge2-freq} are real. Eigenfrequencies $\omega_{7,8}$
in \eqref{gauge1-freq} and \eqref{gauge2-freq} are real, if
$\left(\pd{\alpha^2}{\gamma_{ij}} \right) v_i v_j > 0 $ and
$\left(\pd{\alpha^2}{\gamma_{ij}} + \beta^i\beta^j\right) v_i v_j
> 0 $, respectively. The eigenvectors of $\hat A_8$ in
\eqref{A8} are uniformly linearly independent, for this particular
choice of gauges. For the case of fixed gauges,
$\pd{\alpha^2}{\gamma_{ij}} = 0$, the eigenvalues are still real,
but $\hat A_8$ has only seven linearly independent eigenvectors.
That is, all fixed gauges lead to ill-posed constrained evolution.
The general conditions for well-posedness (strong hyperbolicity)
of the constrained ADM 3+1 formulation with algebraic gauges
\eqref{gauge1} and \eqref{gauge2} therefore are
\labeq{ADM-Condition-1}{
                        \left( \pd{\alpha^2}{\gamma_{ij}} \right) v_i v_j > 0
}
and
\labeq{ADM-Condition-2}{
                        \left( \pd{\alpha^2}{\gamma_{ij}} + \beta^i \beta^j \right) v_i v_j > 0
}
for an arbitrary $v_i$, respectively. Again,
\eqref{ADM-Condition-2} is the result obtained in \cite{11}.

The analysis of matrix $\hat A_8$ in \eqref{A8}  for algebraic
gauges shows that the minimal set \eqref{A8} separates into four
independent subsets each consisting of two equations.
 Among these subsets there are three well-posed subsets corresponding to pairs
 of eigenvalues $\omega_{1,...,6}$. These three subsets
describe propagation of gravitational waves and a gauge wave. We
call these wave subsets. The fourth subset, which corresponds to
eigenvalues $\omega_{7,8}$, describes another two gauge modes. The
solution of the fourth subset depends on solutions of the strongly
hyperbolic wave subsets and the gauge choice. Those four subsets
completely describe the behavior of Einstein's equations
(evolutionary part + constraints) in the high frequency limit. It
is therefore evident that  the posedness of Einstein's equations
depends entirely upon the properties of the gauge.

As a simple illustration, we explicitly present the form of these
subsets for a general metric $\gamma_{ij}$, extrinsic curvature
$K_{ij}$, gauge $\alpha=\alpha(\gamma_{ij},x^n,t)$ and $\beta^k
=\beta^k(x^n,t)$, and propagation of perturbations along $x^1$
coordinate direction, $v_k = (v_1,0,0)$ and
\mbox{$\g^{11}v_1v_1=1$}.
The eigenvalues for this case are: $\{\b^1 v_1-\a,\ \b^1 v_1-\a
,\b^1 v_1+\a
  ,\ \b^1 v_1+\a,$ \\
  $ \ \b^1 v_1,\ \b^1 v_1,\ \b^1 v_1+\a\sqrt{2A^{11}v_1v_1},\ \b^1 v_1-\a\sqrt{2A^{11}v_1v_1} \}$
and the minimal set is explicitly given by the following four
subsets
\labeq{SubSet1}{ {\rm I:}\begin{split}\pd{\delta K_{23}}{t}= &
\alpha v_1\left[-\half\gamma^{11}\pd{\delta
D_{23;1}}{\lambda}+\frac{\beta^1}{\alpha}\pd{\delta
K_{23}}{\lambda}\right],
\\
\pd{\delta D_{23;1}}{t}= & \a
               v_1\left[-2\pd{\delta K_{23}}{\lambda}+\frac{\b^1}{\a}\pd{\delta D_{23;1}}{\lambda}\right]
\end{split}                }
\labeq{Subset2}{{\rm II:} \begin{split}\pd{\delta K_{33}}{t} =
&\alpha v_1\bigg[-\half\gamma^{11}\pd{\delta
D_{33;1}}{\lambda}+\frac{\beta^1}{\alpha}\pd{\delta
K_{33}}{\lambda}\bigg], \\
\pd{\delta D_{33;1}}{t} = &\a
               v_1\bigg[-2\pd{\delta K_{33}}{\lambda}+\frac{\b^1}{\a}\pd{\delta D_{33;1}}{\lambda}\bigg]
\end{split}}
\begin{widetext}
\labeq{SubSet3}{ \hspace{-1cm} {\rm III:} \begin{split}
\pd{\delta D_{12;1}}{t}=& \alpha
               v_1\left\{-\frac{2\left(\gamma^{12}\gamma^{12}\gamma^{13}+\gamma^{11}\gamma^{13}\gamma^{22}-
                     2\gamma^{11}\gamma^{12}\gamma^{23}\right)}
                    {\gamma^{11}(\gamma^{12}\gamma^{12}-\gamma^{11}\gamma^{22})}\pd{\delta K_{23}}{\lambda}
                     -\frac{2\left(\gamma^{12}\gamma^{13}\gamma^{13}-\gamma^{11}\gamma^{12}\gamma^{33}\right)}
                     {\gamma^{11}(\gamma^{12}\gamma^{12}-\gamma^{11}\gamma^{22})}
                     \pd{\delta K_{33}}{\lambda} \right.\\
                     &\qquad\qquad +\frac{\b^1}{\a}\pd{\delta
                     D_{12;1}}{\lambda}\bigg\},
                     \\
\pd{\delta D_{13;1}}{t}=& \alpha
               v_1\left\{\frac{2}{\gamma^{11}}\left[
              \gamma^{12}\pd{\delta K_{23}}{\lambda}+\gamma^{13}\pd{\delta K_{33}}{\lambda}\right]
              +\frac{\b^1}{\a}\pd{\delta
              D_{13;1}}{\lambda}\right\},
\end{split}
}
\labeq{SubSet4}{\hspace{-0.3cm} {\rm IV:}
\begin{split}
\pd{\delta K_{11}}{t}=& \alpha
             v_1\bigg[-A^{11}\pd{\delta D_{11;1}}{\lambda}-2A^{12}\pd{\delta D_{12;1}}{\lambda}-
             2A^{13}\pd{\delta D_{13;1}}{\lambda}-\left(\half\gamma^{22}d^{23}+\gamma^{23}+
             2A^{23}+A^{22}d^{23}\right)\pd{\delta D_{23;1}}{\lambda} \\
             &\qquad\qquad-
             \left.\left(\half\gamma^{22}d^{33}+\half\gamma^{33}
               +A^{33}+A^{22}d^{33}\right)\pd{\delta D_{33;1}}{\lambda}+
                       \frac{\beta^1}{\alpha}\pd{\delta K_{11}}{\lambda}\right], \hspace{2.4cm}\\
    \pd{\delta D_{11;1}}{t}=&
     \a v_1\left[-2\pd{\delta K_{11}}{\lambda}+\frac{\b^1}{\a}\pd{\delta
     D_{11;1}}{\lambda}\right],
\end{split}
}
\\
\end{widetext}
where \labeq{}{\begin{split} A^{ij}=& \pd{\ln\a}{\g_{ij}},\quad
d^{23}=-2\frac{\g^{12}\g^{13}-\g^{11}\g^{23}}{\g^{12}\g^{12}-\g^{11}\g^{22}},
\quad \\
d^{33}=&
-\frac{\g^{13}\g^{13}-\g^{11}\g^{33}}{\g^{12}\g^{12}-\g^{11}\g^{22}}.\end{split}
}
As one can easily see these equations are valid only when
$\g^{12}\g^{12}-\g^{11}\g^{22}\neq 0$. This has been our
assumption to solve the momentum and Hamiltonian constraints for
the derivatives of the dynamical variables which we wanted to
eliminate. Although it may seem that this is not a general result
we point out that the kinematic constraints can always be used to
eliminate 4 of the dynamical variables provided  that certain
conditions are held true. If the condition above is not satisfied
then there will be another set of variables that we will be able
to eliminate and thus obtain the minimally coupled set of partial
differential equations. In this respect we have not lost
generality.

Subsets I, II, and III describe wave propagation and are
well-posed. The first two propagate with the shift plus the
fundamental speed (the lapse function) and the third one with the
shift velocity. Subset IV will be well posed, if the lapse
satisfies $\pd{\ln\alpha}{\gamma_{11}}v_1v_1> 0 \Rightarrow
\pd{\alpha^2}{\gamma_{11}} > 0$, which is a particular case of the
general condition \eqref{ADM-Condition-1}. If $
\pd{\alpha^2}{\gamma_{ij}} = 0$ (fixed gauge), the fourth subset
takes the form

\labeq{SubSet4b}{ {\rm IV:}\begin{split}
 \pd{\delta K_{11}}{t}&= \alpha
v_1\left[-\left(\half\gamma^{22}d^{23}+\gamma^{23}\right)\pd{\delta
D_{23;1}}{\lambda}\right.\\
                    -\bigg(&\half\gamma^{22}d^{33}+\half\gamma^{33}\bigg)\pd{\delta D_{33;1}}{\lambda}+
                       \frac{\beta^1}{\alpha}\pd{\delta K_{11}}{\lambda}\bigg],\\
\pd{\delta D_{11;1}}{t}&=
     \a v_1\left[-2\pd{\delta K_{11}}{\lambda}+\frac{\b^1}{\a}\pd{\delta
     D_{11;1}}{\lambda}\right].
\end{split}
}
This subset is weakly hyperbolic and ill-posed. This can be most
easily seen if we consider a simplest case with $\delta
D_{23;1}=\delta D_{33;1}=0$ and $\beta^1=0$. Then the solution of
\eqref{SubSet4b} will be $\delta K_{11}=\delta K_{11}(\lambda,0)$
and $\delta D_{11;1}(\lambda,t) = \delta D_{11;1}(\lambda,0) +
\left( \pd{\delta K_{11}}{\lambda}(\lambda,0) \right)\,
 t $. The linear growth of  $\delta D_{11;1}$ depends on initial conditions
and may be arbitrarily fast. Since in the high frequency limit we
can treat the gauge functions as constants, the physical
acceleration can be neglected. Therefore, the linear growth of
$\delta D_{11;1}$ physically describes the deformation of a
synchronous reference frame with time, which in a general
non-linear case when the perturbations are not small, leads to the
formation of caustics.

Constrained evolution of  perturbations of all other variables  is
completely determined by the solution of subsets I - IV. As an
example we present the evolution of $\delta K_{13}$ and $\delta
D_{22;1}$ when, for simplicity, the shift vector is zero :
\labeq{SubSet5}{\begin{split} \pd{\delta K_{13}}{t} = & \alpha v_1
\left(\half\gamma^{12}
       \pd{\delta D_{23;1}}{\lambda} + \half\gamma^{13} \pd{\delta D_{33;1}}{\lambda}\right),
  \\     \quad \pd{\delta D_{22;1}}{t} =  & -2 \alpha v_1\left[
                     2\frac{\left(\gamma^{11}\gamma^{23}-\gamma^{12}\gamma^{13}\right)}
                     {(\gamma^{12}\gamma^{12}-\gamma^{11}\gamma^{22})}\pd{\delta K_{23}}{\lambda}\right.\\
                     &\left.+
                     \frac{\left(\gamma^{11}\gamma^{33}-\gamma^{13}\gamma^{13}\right)}
                     {(\gamma^{12}\gamma^{12}-\gamma^{11}\gamma^{22})}\pd{\delta
                     K_{33}}{\lambda}\right].
\end{split}
}
The amplitudes of $\delta K_{13}$ and $\delta D_{22;1}$ will not
grow. Equations for $\delta K_{22}$ and $\delta K_{12}$ are
analogous but more complicated. All other perturbations satisfy
equations
\labeq{}{
           \pd{\delta D_{ij;2}}{t} = \pd{\delta D_{ij;3}}{t} = 0.
}
We found that the behavior described above is similar to that of a
general case of algebraic gauges and any arbitrary direction of
propagation of perturbations.

Examples of algebraic gauges are the widely used gauge
$\alpha=C(x^i)\gamma^{1/2}$ often referred to as the ``harmonic''
gauge \cite{3}, the ``1+log'' gauge $\alpha=1 + \log(\gamma)$
\cite{18}, and the densitized lapse gauge $\alpha=C(x^i)
\gamma^\sigma$ \cite{4}, all depending on the determinant of the
three-metric, $\gamma=det(\gamma_{ij})$. For these gauges,
condition \eqref{ADM-Condition-1} can be written as
 \labeq{DetgCondition}{\pd{\ln \a}{\gamma_{ij}}v_iv_j=\pd{\ln
\a}{\gamma}\pd{\g}{\gamma_{ij}}v_iv_j=\pd{\ln
\a}{\gamma}\g{\gamma^{ij}}v_iv_j=\pd{\ln \a}{\ln\gamma} >0.}
It can be readily seen that both ``harmonic'' and ``1+log'' gauges
satisfy this condition and lead to a well posed constrained
evolution. The densitized lapse will provide a well-posed
constrained evolution only if
\labeq{SigmaCondition}{\sigma >0.}

\subsection{\label{ADM_plus_Dif_Gauges}Well-posedness of ADM with differential gauges}

Similar approach to posedness of 3+1 formulations can be carried
out for more complex gauges involving non-zero shift and general
elliptic, parabolic, or hyperbolic differential gauges. As
examples of elliptic and parabolic differential gauges we will
consider the maximal slicing condition and its parabolic
extension. Although these two gauges are believed to prevent
coordinate singularities, here we demonstrate that both of them
produce weakly hyperbolic minimal sets and thus produce ill-posed
constrained evolution.

The maximal slicing condition \cite{14} is $K=0$, where $K$ is the
trace of the extrinsic curvature. In the following analysis we
need the evolution equations for $K$ and the determinant $\gamma$
of the 3-metric in vacuum. Those can be derived by taking the
traces of \eqref{ADM-gamma} and \eqref{ADM-K} and they are
\labeq{det_evol}{
\partial_t\ln\gamma^{1/2}=-\alpha K+\nabla_i \beta^i}
\labeq{K_evol}{\partial_t
K=-\gamma^{ij}\nabla_i\nabla_j\alpha+\alpha
K_{ij}K^{ij}+\beta^i\nabla_i K }
If the $K=0$ condition is imposed then \eqref{K_evol} results in
the following elliptic differential equation for the lapse
function
\labeq{Elliptic2}{\gamma^{ij}\nabla_i\nabla_j\alpha-\alpha
K_{ij}K^{ij}=0. }
If the Hamiltonian constraint \eqref{ADM-H} is satisfied
\eqref{Elliptic2} yields
\labeq{Elliptic}{\gamma^{ij}\nabla_i\nabla_j\alpha-\alpha R=0. }
In the limit of high frequency perturbations the Ricci scalar
vanishes on the surface of constraints and equation
\eqref{Elliptic} reduces to
\labeq{}{\gamma^{ij}\partial_i\partial_j \delta\alpha=0, } which,
if written along a given direction $v^i$, yields
\labeq{Maximal_problem}{\g^{ij}v_iv_j\pdd{\delta
\alpha}{\lambda}=\pdd{\delta \alpha}{\lambda}=0.}
Then the principal term of $\nabla_i\nabla_j\delta \alpha=
v_iv_j\pdd{\delta \alpha}{\lambda}$ vanishes and therefore the ADM
+ Maximal Slicing equations have the same properties as ADM +
Fixed Gauges, which means that the Cauchy problem for ADM+Maximal
Slicing is ill posed. We should keep in mind that result
\eqref{Maximal_problem} is not only valid for a constrained
evolution, but also for an unconstrained one, since it can be
derived from \eqref{Elliptic2}, too. Equation
\eqref{Maximal_problem} results because of the high frequency
perturbations we are considering here.

Here we ought to resolve the apparent contradiction between our
analysis and the fact that maximal slicing prevents the formation
of coordinate singularities \cite{14}. This can be seen if
\eqref{det_evol} is written as
\labeq{lie}{K=\pounds_{\hat{\bf n}}\ln(\gamma^{1/2}),}
where $\pounds_{\hat{\bf n}}$ denotes the Lie derivative along the
unit normal vector $\hat{{\bf n}}$ to the spacelike hypersurfaces
with 3 metric $\gamma_{ij}$. If we set $K=0$ then from \eqref{lie}
the local volume element $\gamma^{1/2}$ is proper-time independent
and  cannot shrink to zero. This means that a coordinate
singularity cannot be formed.

However, the well-posedness properties of algebraic $K=0$ slicing
condition are different from those of \eqref{Elliptic} with
initial conditions $K=0$. If maximal slicing is imposed by using
$K=0$ at all times by eliminating one of the components, for
example $ K_{11} = - \frac{\gamma^{ij} - \delta^{1i}
\delta^{1j}\gamma^{11}}{\gamma^{11}} K_{ij} $ this reduces the
number of dynamical degrees of freedom in the minimal set
\eqref{A8} from eight to seven \footnote{ We must still use
\eqref{Elliptic} to determine the lapse.}. In this case the
perturbations of the trace of the extrinsic curvature are
identically zero, $\delta K = 0$, and the minimal set of seven
equations is well posed. On the other hand, if we use
\eqref{Elliptic} and initial conditions $K(t=0)=0$ alone, the
minimal set of eight equations is ill-posed because $\delta K$ are
not necessarily zero.  If we write equation \eqref{K_evol} in the
limit of high frequency perturbations, keeping
\eqref{Maximal_problem} in mind, we obtain
\labeq{}{\partial_t \delta K=\beta^i\partial_i \delta K.}
Although this equation is well-posed and hence the perturbations
of $K$ do not grow, we notice that the $K=0$ condition may now be
violated by perturbations of $K$. It is this violation which is
the root to the ill-posedness associated with \eqref{Elliptic}.

Let us illustrate this  with the following example. Consider
planar high frequency perturbations along $x^1$ about Minkowski
spacetime, which means that the unperturbed lapse is $\a=1$ and
the unperturbed shift is $\b^i=0$. In this case, the minimal set
for the differential maximal slicing condition contains
\eqref{SubSet4b}, which for this case can be written as
\labeq{SubSet4b_maximal}{ {\rm IV:}\begin{split}
 \pd{\delta K_{11}}{t}&=0, \\
\pd{\delta D_{11;1}}{t}&=-2v_1\pd{\delta K_{11}}{\lambda}.
\end{split}
}
This subset \eqref{SubSet4b_maximal} is weakly hyperbolic and
hence ill-posed. In addition, if we impose the algebraic maximal
slicing condition at all times, then in the high frequency limit
\labeq{K0}{\pd{\delta K_{11}}{\lambda}=-(\pd{\delta
K_{22}}{\lambda}+\pd{\delta K_{33}}{\lambda}).}
This equation eliminates $\delta K_{11}$ from the minimal set. The
linearized momentum constraints for this particular case also read
\labeq{mom}{\pd{\delta K_{22}}{\lambda}+\pd{\delta
K_{33}}{\lambda}=0.}
Combining \eqref{K0} and \eqref{mom} we obtain $\pd{\delta
K_{11}}{\lambda}=0$ and
\eqref{SubSet4b_maximal} reduces to one equation
\labeq{SubSet4b_maximal_good}{ {\rm IV:} \pd{\delta
D_{11;1}}{t}=0. }
Equation \eqref{SubSet4b_maximal_good} is well posed, and this
eliminates the possibility of formation of coordinate
singularities.

The parabolic extension of maximal slicing \cite{17} has the
following form
 \labeq{Parabolic}{\pd{
 \alpha}{t}=\frac{1}{\epsilon}\left(\gamma^{ij}D_iD_j\alpha-K_{ij}K^{ij}\alpha-c K\right),
} where $\epsilon$ is a positive constant. Since the lower order
term $-K_{ij}K^{ij}\alpha-c K$ does not belong to the principal
part, application of high frequency perturbations along any
arbitrary direction yields
 \labeq{Parabolic2}{\pd{\delta \alpha}{t}=\frac{1}{\epsilon}\pdd{\delta
 \alpha}{\lambda}.
} The linearized evolution equations for
 variables belonging to the minimal set for this particular gauge
  are given by (see \eqref{dKij} and \eqref{dDijk}):
\labeq{dKijParabolic}{ \pd{\delta K_{ij}}{t}  =
\alpha\left(-\frac{1}{\alpha}v_iv_j\pdd{\delta\alpha}{\lambda}+R_{ij}^{1}\right)
                   +\Gamma^k_{ij}v_k\pd{\delta\alpha}{\lambda},
}
\labeq{dDijkParabolic}{ \pd{\delta D_{ij;k}}{t}  = -2\a
                     v_k\pd{\delta
                     K_{ij}}{\lambda}-2K_{ij}v_k\pd{\delta\alpha}{\lambda}.
}
In the high frequency limit we are considering here, the evolution
equation for the lapse \eqref{Parabolic2} is completely decoupled
from \eqref{dKijParabolic} and \eqref{dDijkParabolic} whatsoever.
But, the linearized ``parabolic maximal slicing" is a diffusion
equation, which is well-posed and it dictates that the amplitude
of the perturbation of the lapse function in this case diffuses
out and hence it does not grow.  This means that as time passes
the entire system will asymptotically resemble  the case of fixed
lapse and zero shift, which is a special case of a fixed gauge.
Therefore the constrained evolution of ADM with ``parabolic
maximal slicing" is ill-posed. In appendix \ref{appB} we present a
more rigorous proof of this fact. These results for maximal
slicing and its parabolic extension agree with those obtained in
\cite{11} for the same slicing conditions.

Following Bona {\it et al.} \cite{12} hyperbolic gauges can be
given in general by the following conditions:
\labeq{}{\begin{split}
\ptl_t\a=& -\a^2 Q,  \\
\ptl_t\b^i=& -\a Q^i,
\end{split}
}
where $Q, Q^i$ will be given by either algebraic or differential
equations relating them with other variables of the system and
will be chosen accordingly in order to obtain hyperbolic equations
for the lapse and/or the shift. For such slicings, one has to
modify the ADM equations by defining new variables in order to
obtain the 1st order form. Therefore we define $A_i=\ptl_i\ln\a,\
B_j{}^l=\ptl_j\b^l$ and with these definitions we compute:
$\ptl_i\ptl_j\a=\a (A_i A_j+\ptl_i A_j)$ and
$\ptl_i\ptl_j\b^l=\ptl_i B_j{}^l$. Thus, the linearized principal
part of the ADM equations can be written as
\labeq{ADMPPDif}{
\begin{split}
\pd{\delta \gamma_{ij}}{t}  =& 2
\gamma_{\ell(i}\delta B_{j)}{}^\ell, \\
\pd{\delta K_{ij}}{t}  = & \alpha\left( R_{ij}^{1}+
                   \frac{\beta^k}{\alpha} \pd{\delta K_{ij}}{x^k}-\pd{\delta  A_j}{x^i} \right), \\
\pd{\delta D_{ij;k}}{t}  =& \alpha\left[-2\pd{\delta
K_{ij}}{x^k}+\frac{\beta^{\ell}}{\alpha}\pd{\delta
D_{ij;k}}{x^\ell} \right.\\
& \left.\quad+\frac{1}{\a}\bigg(\g_{li}\pd{ \delta
B_j{}^l}{x^k}+\g_{lj}\pd{ \delta B_i{}^l}{x^k}\bigg) \right].
\end{split}
}
\\
We have introduced twelve new variables and therefore we need
twelve additional evolution equations to describe them. One could
expect that the number of differential equations of the minimal
set would be 8+12=20 equations. However,  we must impose eight new
linear constraints arising due to the introduction of new
variables $B^k_i$ and $A_i$,
\labeq{ShiftCon1}{\pd{B_i{}^k}{x^j}=\pd{B_j{}^k}{x^i} }
and
\labeq{LapseCon1}{\pd{A_i}{x^j}=\pd{A_j}{x^i}. }
For perturbations moving along direction $v_i$ and assuming $v_1
\neq 0$ we get
\labeq{ShiftCon}{\pd{B_i{}^k}{\lambda}=\frac{v_i}{v_1}\pd{B_1{}^k}{\lambda}
} and
\labeq{LapseCon}{\pd{A_i}{\lambda}=\frac{v_i}{v_1}\pd{A_1}{\lambda}.
}
Equations \eqref{ShiftCon} and \eqref{LapseCon} mean that there
are 3 linearly independent $B_i{}^k$ and 1 independent $A_i$.
Therefore the minimal set will in principle consist of only twelve
evolution equations for our dynamical variables.

As an example we will consider the Bona-Masso family of slicing
conditions \cite{16} which we write in terms of the new variables
as
\labeq{}{\ptl_t\ln\a=\b^i A_i-\a f(\a)(K-K_o),}
where $f(a)$ is a strictly positive function of the lapse, K is
the trace of the extrinsic curvature and $K_o=K(t=t_o)$. Then the
principal part of the evolution equations of the perturbations of
the new variables is
\labeq{}{\ptl_t \delta A_i=\b^k\ptl_k \delta A_i-\a
f(\a)\g^{kl}\pd{\delta K_{kl}}{x^i}. } However, because of
constraints \eqref{LapseCon} only one of the above equations will
be a part of the minimal set. Our analysis of the minimal set,
carried out for this gauge, shows that the constrained evolution
is well posed.

As an illustration of the latter, we present the linearized
constrained evolution equations for the zero shift vector case
(also known as the K-driver condition). The minimal set then
consists of only nine partial differential equations (three
equations get eliminated due to $\beta^i=0$) for planar
perturbations of the dynamical variables $K_{11}$, $K_{23}$,
$K_{33}$, $D_{11;1}$,$ D_{12;1}$, $D_{13;1}$, $D_{23;1}$,
$D_{33;1}$, and $A_1$ along the $x^1$-axis, which we group into
the following subsets
\labeq{DSubSet1}{{\rm I:}\quad\begin{split} \pd{\delta K_{23}}{t}=
& -\half\alpha v_1\gamma^{11}\pd{\delta D_{23;1}}{\lambda}, \\
\quad \pd{\delta D_{23;1}}{t}= & -2\a
               v_1\pd{\delta K_{23}}{\lambda}
  \end{split}
               }
\labeq{DSubSet2}{{\rm II:} \begin{split}\quad \pd{\delta
K_{33}}{t}=& -\half\alpha v_1\gamma^{11}\pd{\delta
D_{33;1}}{\lambda},\\ \quad
\pd{\delta D_{33;1}}{t}=& -2\a v_1\pd{\delta K_{33}}{\lambda}
\end{split}}

\begin{widetext}
\labeq{DifSubSet3}{{\rm III:}\quad \begin{split} \pd{\delta
D_{12;1}}{t}=& -2\a v_1
               \left[
                     \frac{\left(\gamma^{12}\gamma^{12}\gamma^{13}+\gamma^{11}\gamma^{13}\gamma^{22}-
                     2\gamma^{11}\gamma^{12}\gamma^{23}\right)}
                     {\gamma^{11}(\gamma^{12}\gamma^{12}-\gamma^{11}\gamma^{22})}
                     \pd{\delta K_{23}}{\lambda}+
                     \frac{\left(\gamma^{12}\gamma^{13}\gamma^{13}-\gamma^{11}\gamma^{12}\gamma^{33}\right)}
                     {\gamma^{11}(\gamma^{12}\gamma^{12}-\gamma^{11}\gamma^{22})}
                     \pd{\delta K_{33}}{\lambda}\right],
               \\
\pd{\delta D_{13;1}}{t}=& \frac{2\a v_1}{\gamma^{11}}\left(
              \gamma^{12}\pd{\delta K_{23}}{\lambda}+\gamma^{13}\pd{\delta
              K_{33}}{\lambda}\right)
              \end{split}
}
\end{widetext}
\labeq{DifSubSet4}{{\rm IV:} \quad
\begin{split}
\pd{\delta K_{11}}{t}=& \ \alpha
v_1\left[-\left(\half\gamma^{22}d^{23}+\gamma^{23}\right)\pd{\delta
D_{23;1}}{\lambda}\right.
  \\ &
                 -      \half\left(\gamma^{22}d^{33}+\gamma^{33}\right)\pd{\delta
                       D_{33;1}}{\lambda}-\pd{A_1}{\lambda}\bigg],
                        \\
\pd{ \delta A_1}{t}= &-\a
                f(\a)v_1\bigg[\g^{11}\pd{\delta K_{11}}{\lambda}
                \\ &
                +2\frac{\g^{12}(\g^{13}\g^{22}-\g^{12}\g^{23})}
                {\gamma^{12}\gamma^{12}-\gamma^{11}\gamma^{22}}\pd{\delta
                K_{23}}{\lambda}
                \\ &
                +
\frac{\g^{13}\g^{13}\g^{22}-\g^{12}\g^{12}
   \g^{33}}{\gamma^{12}\gamma^{12}-\gamma^{11}\gamma^{22}}\pd{\delta
K_{33}}{\lambda}\bigg]
        \end{split}}

        and an equation for the evolution of $\delta D_{11;1}$
\labeq{DifD111}{ \pd{\delta D_{11;1}}{t}= -2v_1\a\pd{\delta
K_{11}}{\lambda}.}

Subsets I and II describe gravitational waves propagating with the
fundamental speed. The eigenvalues which correspond to these two
subsets are $\{-\a,-\a,\a,\a\}$. These are well-posed and
completely decoupled from the rest of the system. The
perturbations which correspond to subset III are completely
determined by the solution of the first two subsets and they do
not grow. Two zero eigenvalues correspond to the third subset and
therefore it corresponds to static modes. The fourth subset is
coupled to the first two, but it is not completely defined by the
solution of I and II. Subset IV is also well posed and it
describes a gauge wave propagating with speed $\a\sqrt{f(\a)}$.
Finally equation \eqref{DifD111}, which describes a static gauge
mode, is completely determined by the solution of the fourth
subset and the perturbation for $\delta D_{11;1}$ does not grow.
It is clear therefore that this system of equations has a
well-posed Cauchy problem. We found exactly the same behavior in
the most general case of perturbations of arbitrary direction. The
Jordan matrix is always diagonal and hence the system is
well-posed. If the shift vector is fixed, but non vanishing, its
presence does not affect the well-posedness of the constrained
system, since the Jordan matrix is still diagonal and the 9 non
zero real eigenvalues are
\labeq{bona-freq}{ \b^iv_i,\b^iv_i,\b^iv_i,\b^iv_i \pm \a,\b^iv_i
\pm \a,\b^iv_i \pm \a\sqrt{f(\a)} }

One can demonstrate, as in the previous section, that the solution
of the entire linearized system can be retrieved once the minimal
set is solved and that its posedness is completely dependent on
the posedness of the minimal set. Thus, we conclude that the
Bona-Masso family of slicing conditions gives rise to a well-posed
constrained ADM evolution.

\section{Analysis of extended 3+1 formulations}

The analysis of the standard ADM 3+1 formulation presented in the
previous section can be applied to other 3+1 formulations of GR.
In this section we will explicitly study two of those
re-formulations of GR, namely the Kidder-Scheel-Teukolsky (KST)
\cite{4} (and all KST-like formulations) and the
Baumgarte-Shapiro-Shibata-Nakamura (BSSN) \cite{3}.

\subsection{ The Kidder-Scheel-Teukolsky formulation of 3+1 GR}

Kidder {\it et al.} \cite{4} suggested a new formulation of
Einstein's equations with a strongly (or even symmetric)
hyperbolic set of evolution equations. They obtained this
formulation by adding terms proportional to the constraint
equations to the RHS of the ADM evolution equations. This does not
change the physics the equations describe but changes the
character of partial differential equations which describe the
free evolution. The modified set they suggested is (using their
notation for this section only)

\labeq{KST}{\begin{split}
\partial_t K_{ij}= & (\ldots)+\gamma N g_{ij}\mathcal{C}+\zeta N g^{ab}\mathcal{C}_{a(ij)b}, \\
\partial_t d_{kij}=& (\ldots)+\eta Ng_{k(i}\mathcal{C}_{j)}+\chi
Ng_{ij}\mathcal{C}_k,
\end{split}
}
\\
where $(\ldots)$ stands for the RHS of the ADM evolution
equations, $N$ is the lapse function, $g_{ab}$ is the 3-metric,
$K_{ij}$ the extrinsic curvature, $d_{kij}$ are the same variables
as our $D_{ij;k}$, $\mathcal{C}$ and $\mathcal{C}_i$ are the
hamiltonian and momentum constraints, respectively and
\labeq{}{\mathcal{C}_{klij}\equiv\partial_{[k}d_{l]ij}=0. }
Finally, $\{\gamma,\zeta,\eta,\chi\}$ are arbitrary constants.

The system was closed with the densitized lapse \labeq{}{
Q\equiv\ln(Ng^{-\sigma}), } where $Q$ is a function independent of
the dynamical fields, $g$ is the determinant of the three-metric
and $\sigma$ is the densitization parameter, and was found that
$\sigma > 0$ is essential for obtaining a well-posed set of
evolution equations. This is exactly what we obtain by our
analysis \eqref{SigmaCondition}, without adding constraints to the
ADM equations, but explicitly imposing them.

If we apply our constrained perturbation analysis to the KST
formulation we will cancel the added constraints on the RHS of
their formulation. As result the KST formulation has exactly the
same  analysis as the ADM formulation.

According to \cite{4} any transformation of dynamical variables
does not change the hyperbolic classification of a set of PDEs, if
this transformation satisfies the following conditions
\begin{enumerate}
\item The transformation is linear in all dynamical variables
except possible the metric

\item The transformation is invertible

\item Time and space derivatives of the metric can be written as a
sum of only the non-principal terms.
\end{enumerate}
Their redefinition of variables and the introduction of
``kinematical'' ones is a transformation which satisfies the
aforementioned criteria and thus it does not affect the hyperbolic
properties of the set of evolution equations. Hence, the
constrained perturbation analysis of the KST formulation and all
KST-like formulations (i.e. all those formulations which are
derived by addition of constraints to the RHS of ADM and perhaps a
transformation of variables with the aforementioned properties) is
equivalent to that of the ADM equations.

This argument may also be used to conclude that any 3+1 system
directly obtained from ADM using the above transformation will be
equivalent to ADM.

\subsection{ The Baumgarte-Shapiro-Shibata-Nakamura formulation of 3+1 GR \label{BSSNFormulation}}

The BSSN formulation was initially introduced by Nakamura {\it et
al.} \cite{3}, then modified by Shibata and Nakamura \cite{3}, and
it was later reintroduced slightly modified by T. Baumgarte and S.
Shapiro \cite{3}. Before we proceed with our analysis let us
review the BSSN formulation first. In what follows, we will use
the notation introduced by Baumgarte and Shapiro \cite{3}.

\subsubsection{Basic variables and equations}

The fundamental dynamical variables of BSSN are
($\varphi,\tilde{\gamma}_{ij}$,$K$,$\tilde{A}_{ij}$,$\tilde{\Gamma}^i$)
instead of ($\gamma_{ij}$,$K_{ij}$), where
\labeq{BSSN-variables}{\begin{split} \varphi =& (1/12)\log ({\rm
det}\gamma_{ij}),
\\
\tilde{\gamma}_{ij} = & e^{-4\varphi}\gamma_{ij},
\\
K  =& \gamma^{ij}K_{ij},
\\
\tilde{A}_{ij} =& e^{-4\varphi}(K_{ij} - (1/3)\gamma_{ij}K),
\\
\tilde{\Gamma}^i =& \tilde{\Gamma}^i_{jk}\tilde{\gamma}^{jk}.
\end{split}}

The ``connection" symbols $\tilde{\Gamma}^i_{jk}$ are the
Christoffell symbols associated with the conformal three-metric
$\tilde{\gamma}_{ij}$. In the BSSN formulation, the Ricci
curvature tensor is calculated as
\labeq{BSricci}{\begin{split} R^{BSSN}_{ij} =&
R^\varphi_{ij}+\tilde R_{ij},
\\
R^\varphi_{ij} =& -2\tilde{D}_i\tilde{D}_j\varphi
-2\tilde{\gamma}_{ij}\tilde{D}^k\tilde{D}_k\varphi
\\
& +4(\tilde{D}_i\varphi)(\tilde{D}_j\varphi)
-4\tilde{\gamma}_{ij}(\tilde{D}^k\varphi)(\tilde{D}_k\varphi),
\\
\tilde{R}_{ij} =&
-(1/2)\tilde{\gamma}^{lk}\partial_{l}\partial_{k}\tilde{\gamma}_{ij}
+\tilde{\gamma}_{k(i}\partial_{j)}\tilde{\Gamma}^k \\
& +\tilde{\Gamma}^k\tilde{\Gamma}_{(ij)k}
+2\tilde{\gamma}^{lm}\tilde{\Gamma}^k_{l(i}\tilde{\Gamma}_{j)km}
+\tilde{\gamma}^{lm}\tilde{\Gamma}^k_{im}\tilde{\Gamma}_{klj},
\end{split}}
where $\tilde{D}_i$ is the covariant derivative associated with
$\tilde{\gamma}_{ij}$.

The evolution equations for these dynamical variables are
\labeq{BSSN-Evol}{\begin{split}
\partial_t \varphi =& -(1/6)\alpha
K+(1/6)\beta^i(\partial_i\varphi)+(\partial_i\beta^i),
\\
\partial_t \tilde{\gamma}_{ij}=& -2\alpha\tilde{A}_{ij} +\tilde{\gamma}_{ik}(\partial_j\beta^k)
+\tilde{\gamma}_{jk}(\partial_i\beta^k)
\\ &
-(2/3)\tilde{\gamma}_{ij}(\partial_k\beta^k)
+\beta^k(\partial_k\tilde{\gamma}_{ij}),
\\
\partial_t K=& -D^iD_i\alpha +\alpha \tilde{A}_{ij}\tilde{A}^{ij}
+(1/3) \alpha K^2 +\beta^i (\partial_i K),
\\
\partial_t \tilde{A}_{ij}=& -e^{-4\varphi}(D_iD_j\alpha)^{TF} +e^{-4\varphi} \alpha
(R^{BSSN}_{ij})^{TF}
\\ &
+\alpha K\tilde{A}_{ij} -2\alpha \tilde{A}_{ik}\tilde{A}^k{}_j
+(\partial_i\beta^k)\tilde{A}_{kj}
\\&
+(\partial_j\beta^k)\tilde{A}_{ki}
-(2/3)(\partial_k\beta^k)\tilde{A}_{ij} +\beta^k(\partial_k
\tilde{A}_{ij}),
\\
\partial_t \tilde{\Gamma}^i =& -2(\partial_j\alpha)\tilde{A}^{ij} +2\alpha
\big(\tilde{\Gamma}^i_{jk}\tilde{A}^{kj}
-(2/3)\tilde{\gamma}^{ij}(\partial_j K)
\\ &
+6\tilde{A}^{ij}(\partial_j\varphi) \big)  -\partial_j \big(
\beta^k(\partial_k\tilde{\gamma}^{ij})
-\tilde{\gamma}^{kj}(\partial_k\beta^{i})
\\&
-\tilde{\gamma}^{ki}(\partial_k\beta^{j})
+(2/3)\tilde{\gamma}^{ij}(\partial_k\beta^k) \big),
\end{split}
}
\\
where all trace free (TF) two index quantities $T_{ij}$ are given
by
\labeq{}{T_{ij}^{(TF)}=T_{ij}-\frac{1}{3}\g_{ij}T, \quad
T=\g^{kl}T_{kl}.}

The constraint equations are
\begin{eqnarray}{\cal
H}^{BSSN} &=& R^{BSSN}+K^2-K_{ij}K^{ij}=0,\label{BSSNconstraintH}
\\
{\cal M}^{BSSN}_i & =& {\cal M}^{ADM}_i=0,\label{BSSNconstraintM}
\\
{\cal G}^i & =& \tilde{\Gamma}^i-\tilde{\gamma}^{jk}
\tilde{\Gamma}^i_{jk}=0 \label{GammaConstraint},
\\
{\cal A}& =&\tilde{A}_{ij}\tilde{\gamma}^{ij}=0
\label{Tracelessness},
\\
{\cal S} & =& \tilde{\gamma}-1=0.\label{determinant}
\end{eqnarray}
${\cal H}^{BSSN}$ and ${\cal M}^{BSSN}_i$ are the Hamiltonian and
momentum constraints (the kinematic constraints) respectively,
while the latter three are algebraic constraints due to the
requirements of BSSN formulation.

\subsubsection{Linearized equations of BSSN}

The formulation, as given above, is first order in time and second
order in spatial derivatives. The second order derivatives occur
in the evolution equations for the conformal traceless extrinsic
curvature.   In order to apply our analysis we need the first
order form. Therefore, we define new variables
\labeq{}{\partial_m\varphi=\phi_m  \mbox{\ \ \ and \ \ }\partial_m
\tilde{\gamma_{ij}}=\tilde{D}_{ij;m},}
where ``;" does not imply a covariant derivative, but it separates
indices of different nature. In terms of those variables the BSSN
equations linearized with respect to a certain background
spacetime solution (not necessarily a Minkowski spacetime)
\labeq{BSSNBackground}{ \varphi, \quad \tg_{ij}, \quad
\ta_{ij},\quad K, \quad \tilde{\Gamma}^i, \quad \td_{ij;k},\quad
\phi_i} can be written as
\labeq{Linearized-BSSN-Evol}{\begin{split}
\partial_t \delta \varphi =& \ptl_i\delta \beta^i,
\\
\partial_t \delta \tilde{\gamma}_{ij}=& \tilde{\gamma}_{ik}(\partial_j\delta \beta^k)
+\tilde{\gamma}_{jk}(\partial_i\delta
\beta^k)-(2/3)\tilde{\gamma}_{ij}(\partial_k\delta \beta^k),
\\
\partial_t \delta K=& -e^{-4\varphi}\tg^{ik}\ptl_i\ptl_k\delta \alpha +\beta^i (\partial_i \delta K),
\\
\partial_t \delta \ta_{ij}=& -e^{-4\varphi}(\ptl_i\ptl_j\delta \alpha)^{TF}
+e^{-4\varphi} \alpha (\delta R^{BSSN}_{ij})^{TF}
\\ &
+(\partial_i\delta \beta^k)\tilde{A}_{kj} +(\partial_j\delta
\beta^k)\tilde{A}_{ki}
\\&
-(2/3)(\partial_k\delta \beta^k)\tilde{A}_{ij} +\beta^k(\partial_k
\delta \ta_{ij}),
\\
\partial_t \delta \tilde{\Gamma}^i =& -2(\partial_j\delta \alpha)\tilde{A}^{ij}
-(4/3)\a\tilde{\gamma}^{ij}(\partial_j \delta K)
\\&
-\beta^k\partial_j (\delta \td_{ij;k})
+\tilde{\gamma}^{kj}(\partial_j\partial_k\delta \beta^{i}) \\&
+\tilde{\gamma}^{ki}(\partial_j\partial_k\delta \beta^{j})
-(2/3)\tilde{\gamma}^{ij}(\partial_j\partial_k\delta \beta^k),
\end{split}
}
\labeq{D phi Evol}{\begin{split}
\partial_t\delta\tilde{D}_{ij;k}=& -2\a\pd{\delta\tilde{A}_{ij}}{x^k}+\beta^\ell\pd{\delta\td_{ij;\ell}}{x^k}
                            +\tg_{i\ell}\pddd{\delta\b^\ell}{x^j}{x^k}
                            \\ &
                            +\tg_{j\ell}\pddd{\delta\b^\ell}{x^i}{x^k}
                            -\frac{2}{3}\tg_{ij}\pddd{\delta\b^\ell}{x^\ell}{x^k}
                            -2\ta_{ij}\pd{\delta\a}{x^k}
                            \\&
                            +\td_{ij;\ell}\pd{\delta\b^\ell}{x^k}
                            +2(\ptl_{(i}\delta\beta^\ell)\td_{j)\ell;k}
                            \\ &
                            -\frac{2}{3}(\ptl_\ell\delta\b^\ell)\td_{ij;k},
                            \\
\partial_t\phi_k=& -\frac{1}{6}\pd{\delta K}{x^k}+\frac{1}{6}\b^{i}\pd{\delta\phi_i}{x^k}+\pddd{\delta\b^i}{x^i}{x^k}
                  \\ &
                  -\frac{1}{6}K\pd{\delta\a}{x^k}+\frac{1}{6}\phi_i\pd{\delta\b^i}{x^k},
\end{split}
} %
where \labeq{BSSNPerturb}{ \delta \varphi, \quad \delta \tg_{ij},
\quad \delta \ta_{ij},\quad \delta K, \quad \delta
\tilde{\Gamma}^i, \quad \delta \td_{ij;k},\quad \delta \phi_i}
are small amplitude and high frequency perturbations of
\eqref{BSSNBackground} and
\labeq{Ricci-Perturb}{\begin{split} \delta R^{BSSN}_{ij} =&
-2\partial_j\delta \phi_i
-2\tilde{\gamma}_{ij}\tilde{\gamma}^{\ell k}\partial_\ell\delta
\phi_k-\half\tilde{\gamma}^{lk}\partial_{l}\delta \tilde{D}_{ij;k}
\\ &
+\half\left(\tilde{\gamma}_{ki}\partial_{j}\delta \tilde{\Gamma}^k
+\tilde{\gamma}_{kj}\partial_{i}\delta \tilde{\Gamma}^k\right)
\end{split}
}
In the context of the first order formulation, since the evolution
of the conformal 3-metric is decoupled from the evolution of the
rest of the system (just like in ADM), then the derivative of
constraint \eqref{determinant} provides a constraint for the
$\tilde D_{ij;k}$ variables, which has to be taken into
consideration and is
\labeq{Det}{\tilde\gamma^{ij}\tilde D_{ij;k}=0}
Then, the linearized constraint equations in new variables are
\labeq{Lin-BSSNconstraintH}{ \delta R^{BSSN}=\g^{ij}\delta
R^{BSSN}_{ij}=0,}
\labeq{Lin-BSSNconstraintM}{ \tg^{kl}\pd{\delta
\ta_{ki}}{x^l}-\frac{2}{3}\pd{\delta K}{x^i}=0,}
\labeq{Lin-GammaConstraint}{ \pd{\delta
\tilde{\Gamma}^i}{x^s}=\tg^{ik}\tg^{jm}\pd{\delta\td_{km;j}}{x^s},}
\labeq{Lin-Tracelessness}{\tg^{ij}\pd{\delta\ta_{ij}}{x^k}=0,}
\labeq{Lin-Det}{\tg^{ij}\pd{\delta\td_{ij;k}}{x^s}=0.}
The introduction of additional variables implies the introduction
of new linear constraint equations which for perturbations of
\eqref{BSSNBackground} can be written as follows
\labeq{D phi Constraints}{\partial_m
\delta\tilde{D}_{ij;k}=\partial_k \delta\tilde{D}_{ij;m} \mbox{ \
\ \ \ and \ \ \ }\partial_m \delta\phi_k=\partial_k \delta\phi_m}
and
\labeq{phi gamma Constraints}{\partial_m
\delta\tilde{\gamma}_{ij}=0 \mbox{ \ \ \ \ and \ \ \ }\partial_m
\delta\varphi=0.}

Finally equations \eqref{Linearized-BSSN-Evol}-\eqref{D phi Evol}
have to be supplemented with the linearized gauge equations
\eqref{Gauge}.

\subsubsection{Analysis of well posedness of BSSN with fixed
and algebraic gauges}

Constraints \eqref{D phi Constraints} dictate that there is only
one independent $\phi_k$ and 6 independent $\td_{ij;k}$. Equation
\eqref{phi gamma Constraints} tells us (just like in the case of
the ADM formulation) that the evolution of $\delta\varphi$ and
$\delta \tilde\gamma_{ij}$ is decoupled from the evolution of the
perturbations of the remaining dynamical variables. For BSSN the
Hamiltonian constraint can always be solved for the derivative of
that $\phi_k$ involved in it, thus eliminating $\phi_k$ from the
minimal set. The momentum constraints can be solved for the
spatial derivatives of two $\ta_{ij}$'s and the spatial derivative
of $K$. Constraint \eqref{Tracelessness} can be used for the
elimination of one of the components of $\ta_{ij}$. Finally,
constraints \eqref{Lin-GammaConstraint} completely eliminate the
$\tilde{\Gamma}^i$ variables and \eqref{Lin-Det} can eliminate one
more of the $\td_{ij;k}$ variables. This means that fully imposing
the linearized constraint equations results in a set of linear
PDEs for three $\ta_{ij}$ and five $\td_{ij;k}$, the
well-posedness of which will determine the well-posedness of the
entire linearized system \eqref{Linearized-BSSN-Evol}-\eqref{D phi
Evol}.

In this section we mainly analyze gauges for which the lapse
function is dependent on the coordinates and the dynamical
variables and the shift vector is function of only the
coordinates. With this in mind, the linearized principal part of
the set of variables which will be part of the  BSSN minimal set
can be written as follows
\labeq{BSSN Alg Evol}{\begin{split}
\partial_t\delta \ta_{ij}=&-e^{-4\varphi}(\ptl_i\ptl_j\delta \a-\frac{1}{3}\g_{ij}\g^{k\ell}\ptl_k\ptl_\ell\delta \a)
         \\ &
         + e^{-4\varphi}\a\delta R_{ij}^{BSSN}+\b^k\ptl_k
          \delta \ta_{ij}, \\
\partial_t\delta \tilde{D}_{ij;k}=&
-2\a\ptl_k\delta \tilde{A}_{ij}+\beta^\ell\ptl_\ell\delta
\td_{ij;k}.
\end{split}
}   To obtain the minimal set one has to fully impose constraints
\eqref{Lin-BSSNconstraintH}- \eqref{Lin-Det} on equations
\eqref{BSSN Alg Evol}.

The equations for the constrained evolution of BSSN are extremely
complicated. For a general background solution, we were able to
carry out an analytic analysis of BSSN with fixed gauges only.
Similarly to ADM, we found that the constrained evolution of BSSN
with fixed gauges is ill-posed.

For algebraic gauges, the simplest case possible is perturbations
about a flat space $\g_{ij}=\delta_{ij}$. If we define
$\Delta^{ij}=\pd{\ln\a}{\tg_{ij}}$ and $\Delta=\pd{\ln\a}{\phi}$
our analysis shows that the minimal set has eigenvalues
$\{\b^iv_i,\b^iv_i,\b^iv_i \pm \a,\b^iv_i \pm \a,\b^iv_i
\pm\frac{\sqrt{\td}}{\sqrt{6}}\}$, where
\labeq{BSCondition1}{ \td=\Delta+
                               12\Delta^{ij}v_iv_j-4\Delta^{km}\delta_{km}>0
}
is the necessary condition for all eigenvalues to be real. Since
the Jordan matrix for this case is diagonal the BSSN constrained
evolution will be well-posed if \eqref{BSCondition1} is satisfied.
As in the ADM case one could show that solving the minimal set of
BSSN is adequate to obtain the solution of the entire linearized
system.

As an illustration we present a set of constrained evolution
equations for perturbations of variables \labeq{}{\begin{split}
\delta \ta_{11}&,\ \delta \ta_{22},\ \delta \ta_{23},\ \delta
\td_{11;1},\\
\delta \td_{12;1}&,\ \delta \td_{13;1},\ \delta \td_{22;1},\
\delta \td_{23;1} \end{split} } propagating along the $x^1$
direction and perturbed about flat space.

\labeq{BSubSet1}{{\rm I:} \quad \begin{split}  \pd{\delta
\ta_{23}}{t}=& -\half\a\pd{\delta
\td_{23;1}}{\lambda}+\b^1\pd{\delta \ta_{23}}{\lambda},
\\
\pd{\delta \td_{23;1}}{t}= & -2\a\pd{\delta
\ta_{23}}{\lambda}+\b^1\pd{\delta \td_{23;1}}{\lambda},
\end{split}
}

\labeq{BSubSet2}{{\rm II:} \quad \begin{split}
 \pd{\delta \td_{12;1}}{t}= & \b^1\pd{\delta \td_{12;1}}{\lambda},
\\
 \pd{\delta \td_{13;1}}{t}= & \b^1\pd{\delta \td_{13;1}}{\lambda},
\end{split}
 }

\labeq{BSubSet3}{{\rm III:} \begin{split} \pd{\delta \ta_{11}}{t}=
& \a\bigg\{-\frac{2}{3}\left[(\Delta^{11}-\Delta^{33}+
\frac{\Delta}{8})\pd{\delta \td_{11;1}}{\lambda}\right.
                 \\ &
                 \qquad+2\Delta^{12}\pd{\delta \td_{12;1}}{\lambda}+ 2\Delta^{13}\pd{\delta \td_{13;1}}{\lambda}
                 \\ &
                 \qquad +(\Delta^{22}-\Delta^{33})\pd{\delta \td_{22;1}}{\lambda}
                  \\ &
                  \ \qquad +2\Delta^{23}\pd{\delta \td_{23;1}}{\lambda}
                 \bigg]+\frac{\b^1}{\a}\pd{\delta \ta_{11}}{\lambda}
                 \bigg\}, \\
\pd{\delta \ta_{22}}{t}= &
\a\bigg\{\frac{1}{3}\left[(\Delta^{11}-\Delta^{33}+
                    \frac{\Delta}{8}-\frac{3}{4})\pd{\delta \td_{11;1}}{\lambda}\right.
                 \\ & \qquad +2\Delta^{12}\pd{\delta \td_{12;1}}{\lambda}+ 2\Delta^{13}\pd{\delta \td_{13;1}}{\lambda}
                 \\ & \qquad+
                 (\Delta^{22}-\Delta^{33}-\frac{3}{2})\pd{\delta\td_{22;1}}{\lambda}
                 \\ & \qquad +2\Delta^{23}\pd{\delta \td_{23;1}}{\lambda}
                 \bigg]+\frac{\b^1}{\a}\pd{\delta \ta_{22}}{\lambda}
                 \bigg\}, \\
 \pd{\delta \td_{11;1}}{t}= & -2\a\pd{\delta \ta_{11}}{\lambda}+\b^1\pd{\delta \td_{11;1}}{\lambda}, \\
 \pd{\delta \td_{22;1}}{t}= & -2\a\pd{\delta \ta_{22}}{\lambda}+\b^1\pd{\delta \td_{22;1}}{\lambda}. \\
                   \end{split}
                 }

Subset I corresponds to a gravitational wave propagating with the
shift+fundamental speed along $x^1$. It is decoupled from the
other subsets and it is well-posed. Subset II describes
propagation of two waves, which travel with the shift vector
speed. This subset just like the first one is decoupled from the
rest of the system. It is a wave subset and therefore it is
well-posed. Subset III is coupled to the first two subsets and
hence its solution depends on the solutions of I and II. At a
first glance it may seem as a contradiction that there are 3
subsets when for ADM we had 4. However, subset III consists of two
independent. One of them describes a gravitational wave travelling
with the shift+fundamental speed and the other is a gauge wave
travelling with speed $\b^1 \pm\frac{\sqrt{\td}}{\sqrt{6}}$, if
$\td>0$, where for this case
\labeq{BSCondition2}{\td=\Delta-4(\Delta^{11}+\Delta^{22}+\Delta^{33})+12\Delta^{11},
}
which is a special case of \eqref{BSCondition1}.

The constrained evolution of BSSN with fixed gauges, that is,
$\Delta=\Delta^{ij}=0$ has the same subsets, but III is now
decoupled from I and II:

\labeq{BSubSet3b}{{\rm III:}\quad \begin{split} \pd{\delta
\ta_{11}}{t}= & \b^1\pd{\delta \ta_{11}}{\lambda},
 \\
\pd{\delta \ta_{22}}{t}= & -\frac{\a}{4}\pd{\delta
\td_{11;1}}{\lambda}
                   -\frac{\a}{2}\pd{\delta\td_{22;1}}{\lambda}+\b^1\pd{\delta
                   \ta_{22}}{\lambda},
\\
 \pd{\delta \td_{11;1}}{t}= & -2\a\pd{\delta \ta_{11}}{\lambda}+\b^1\pd{\delta \td_{11;1}}{\lambda}, \\
 \pd{\delta \td_{22;1}}{t}= & -2\a\pd{\delta \ta_{22}}{\lambda}+\b^1\pd{\delta \td_{22;1}}{\lambda}. \\
                   \end{split}
                 }
\\
The four eigenfrequencies of this system are $\{\b^1,\b^1,\b^1 \pm
\a\}$, which are all real. However, the Jordan matrix is not
diagonal, which implies that the principal matrix of
\eqref{BSubSet3b} does not have a complete set of eigenvectors,
hence the constrained evolution is weakly hyperbolic, that is,
ill-posed. To assign a physical meaning to weak hyperbolicity it
is instructive to consider the case of vanishing shift. Then one
can easily see that subset III breaks into two subsets. The
ill-posed subset is:

\labeq{BSIllposed}{ \pd{\delta \ta_{11}}{t}=  0,\quad
 \pd{\delta \td_{11;1}}{t}=  -2\a\pd{\delta \ta_{11}}{\lambda}
                 }
\\
and its solution is $\delta \ta_{11} = \delta\ta_{11}(\lambda, t=0
)$, and $\delta \td_{11;1}(\lambda,t) = \delta
\td_{11;1}(\lambda,0) + \left( \pd{\delta
\ta_{11}}{\lambda}(\lambda,0) \right)\,t $. The linear growth of
$\delta \td_{11;1}$ depends on initial conditions and may be
arbitrarily fast. As in the ADM formulation, physically the linear
growth of $\delta \td_{11;1}$ describes the inertial deformation
of a synchronous reference frame with time which, in general
non-linear case when perturbations are not small leads, to the
formation of caustics.

A special class of algebraic gauges is that with the lapse
function depending on the determinant of the 3-metric, that is,
$\alpha=\alpha(\gamma)$, which in the BSSN formulation obtains the
form $\alpha=\alpha(e^{12\varphi})$, and therefore the lapse does
not depend on the conformal 3-metric, but on the conformal factor
only. Thus the strong hyperbolicity condition \eqref{BSCondition1}
reduces to $\td=\Delta>0$. For ``harmonic'' slicing
$\alpha=c(x^i)e^{6\varphi}$ and   ``$1+\log$'' slicing
$\alpha=1+12\varphi$ already discussed in the previous section it
is easy to show that they produce a well-posed constrained BSSN
formulation. Similarly, for a densitized lapse $ Q=\ln(\a
e^{-12\sigma\varphi})$, we find that the requirement for
well-posedness $\td=\Delta=12\sigma\alpha>0$ yields a necessary
condition for $\sigma$, which is of course the same as
\eqref{SigmaCondition}.

To conclude this sub-section we note that the results of the
analysis presented above do not depend on the order of
linearization and enforcement of algebraic constraints of BSSN.
Instead of linearizing the unconstrained BSSN first, one could
have chosen to reduce the number of variables of BSSN, by
eliminating as many variables as there are algebraic constraints
and then linearize the reduced system. It is straightforward to
check that both ways lead exactly to the same minimal sets.

\subsection{BSSN and Differential Gauges}

One can perform the same constrained perturbation analysis for the
BSSN formulation in conjunction with the differential gauges
considered in the ADM analysis above. However, the numerical
relativity community has recently been resorting to non-trivial
shift conditions, for example \cite{22,23}, as an attempt to
accurately evolve black hole binaries. Therefore, instead of
analyzing the same gauges as in the ADM analysis above, here we
focus on the elliptic ``Gamma freezing" condition \cite{22}, which
was formulated for the BSSN formulation
 \labeq{Gamma Freezing}{\partial_t \tilde \Gamma^i=0
\Longleftrightarrow
\partial_j\partial_t\tilde \gamma^{ij}=0.}
This condition obviously "freezes" the evolution of the $\Gamma^i$
variables, hence the name. By use of \eqref{Gamma Freezing} and
the evolution equations of $\Gamma^i$, from \eqref{BSSN-Evol}, one
obtains the following elliptic equations which the shift vector
has to satisfy.
\labeq{DifGammaFreeze}{\begin{split} &
-2(\partial_j\alpha)\tilde{A}^{ij} +2\alpha
\big(\tilde{\Gamma}^i_{jk}\tilde{A}^{kj}
-(2/3)\tilde{\gamma}^{ij}(\partial_j K)
\\ &
+6\tilde{A}^{ij}(\partial_j\varphi) \big)  -\partial_j \big(
\beta^k(\partial_k\tilde{\gamma}^{ij})
-\tilde{\gamma}^{kj}(\partial_k\beta^{i})
\\&
-\tilde{\gamma}^{ki}(\partial_k\beta^{j})
+(2/3)\tilde{\gamma}^{ij}(\partial_k\beta^k) \big)=0.
 \end{split} }
Here we consider only a 1D perturbation approach about flat space
to show that this gauge is good at least in the case considered.

To reduce the system to first order form we define the derivatives
of the shift vector as new variables $B_{j}{}^{l}=\partial_j
\beta^l$. The derivatives of the shift satisfy \eqref{ShiftCon1}
as in the ADM analysis. Of course the introduction of those 9 new
variables leads to the introduction of 9 new constraints which
have to be fully imposed. In the high frequency limit of small
amplitude perturbations those constraints read $\partial_j
\delta\beta^l=0$ together with \eqref{ShiftCon}. In terms of the
new variables the linearized principal part of
\eqref{DifGammaFreeze} becomes
\labeq{linDifGammaFreeze}{
 \tilde{\gamma}^{kj}\partial_j\delta B_k{}^{i}
 +\frac{1}{3}\tilde{\gamma}^{ij}\partial_k \delta B_j{}^{k}
 -2\alpha\frac{2}{3} \tilde{\gamma}^{ij} \partial_j \delta
 K+\beta^j \partial_j \delta \tilde\Gamma^i=0.
}
Equation \eqref{linDifGammaFreeze} can be treated as 3 more
constraints in our approach, allowing us hence to eliminate the
three independent perturbations $\delta B_i{}^{j}$. Thus, after
imposing all available constraints the minimal set consists of 8
equations, as it was expected. To simplify the analysis further we
study this shift condition in conjunction with a lapse function of
the form $\alpha=\alpha(\varphi)$, that is the lapse depends only
on the determinant of the 3 metric. This simplification results in
the evolution equations of $\delta\tilde A_{ij}$ in
\eqref{BSubSet1}-\eqref{BSubSet3} being the same, but where
$\Delta^{ij}=0$, so we will not write them here. However, the
evolution equations of $\delta\tilde D_{ij;1}$ in
\eqref{BSubSet1}-\eqref{BSubSet3} are different, since we have to
deal with the shift terms now. Those equations change as it is
dictated by the linearized evolution equations of $\delta\tilde
D_{ij;k}$ in \eqref{D phi Evol}, which in new variables read
\labeq{DEvolChange}{\partial_t \delta\tilde
D_{ij;1}=(\ldots)+\tg_{i\ell}\pd{\delta B_1{}^\ell}{x^j}
                            +\tg_{j\ell}\pd{\delta B_1{}^\ell}{x^i}
                            -\frac{2}{3}\tg_{ij}\pd{\delta
                            B_1{}^\ell}{x^\ell},
                            }
Where $(\ldots)$ stands for the RHS of the evolution equations of
$\delta D_{ij;1}$ in \eqref{BSubSet1}-\eqref{BSubSet3}. All other
terms in \eqref{D phi Evol} contribute to the low-order part only.
If one imposes all possible constraints then the evolution
equations of $\delta D_{ij;1}$ yield
\labeq{DEvolgammafreeze}{\begin{split}
\partial_t\tilde D_{11;1}= & 0 \\
\partial_t\tilde D_{12;1}= & 0 \\
\partial_t\tilde D_{13;1}= & 0 \\
\partial_t\tilde D_{22;1}= & (\ldots)+\half\beta^1\pd{\tilde D_{11;1}}{\lambda}+\alpha\pd{\tilde A_{11}}{\lambda} \\
\partial_t\tilde D_{23;1}= & (...)
\end{split} }
The Jordan decomposition of the resulting system shows that the
Jordan matrix is diagonal with the following eigenvalues on the
diagonal
$\{0,0,0,\beta^1,\beta^1+\alpha,\beta^1-\alpha,\beta^1+\alpha,\beta^1-\alpha\}$.
All eigenvalues are real and hence the system is well behaved.
Surprisingly, the eigenvalues which correspond to the functional
form of the lapse, that is, $\pm \alpha \Delta/6$ are missing.
Therefore, the "$\Gamma$-freezing" condition gives rise to a well
behaved constrained 1D evolution, even if the lapse chosen is
fixed or calculated by the maximal slicing condition, because in
1D perturbations about flat space this shift condition eliminates
the gauge waves which correspond to the algebraic lapse considered
when the shift vector is fixed. This is what one truly obtains, if
one carries out the analysis of the "$\Gamma$-freezing" shift in
conjunction with a fixed lapse or maximal slicing.

Those results constitute a good indication
of the well-posedness of the constrained BSSN evolution with this
gauge condition. However, a definite answer requires  a
complete analysis, that is, consideration of planar perturbations about an
arbitrary spacetime. This is a very complicated task.  We will address this in a future paper along with the other
popular shift conditions, the parabolic and hyperbolic
"$\Gamma$-Driver" conditions \cite{22} in conjunction with the
Bona-Maso family of slicing conditions.

\subsection{Equivalence of conditions for well-posedness of constrained evolution of ADM and BSSN}

We will now demonstrate that, in the limit of high frequency
perturbations about flat space, the ADM condition
\eqref{ADM-Condition-1} which can be written as
\labeq{}{A=\pd{\ln\alpha}{\gamma_{ij}}v_iv_j>0}
 is equivalent
to that of BSSN, equation \eqref{BSCondition1}. We remind that
\labeq{}{\tilde{\gamma}_{ij}=e^{-4\varphi}\gamma_{ij} \quad
\mbox{and} \quad \varphi=\frac{1}{12}\ln|\gamma|.}
 Now using that $d\gamma=\gamma\gamma^{ij}d\gamma_{ij}$,
we obtain
\labeq{}{\pd{\varphi}{\gamma_{ij}}=\frac{1}{12}\gamma^{ij}} and
\labeq{}{\pd{\tilde{\gamma}_{km}}{\gamma_{ij}}=e^{-4\varphi}\delta^i_k\delta^j_m-
         \gamma_{km}4e^{-4\varphi}\frac{1}{12}\gamma^{ij}.}
         Since,
\labeq{}{\pd{\ln\alpha}{\gamma_{ij}}=\pd{\ln\alpha}{\varphi}\pd{\varphi}{\gamma_{ij}}+
                                  \pd{\ln\alpha}{\tilde{\gamma_{km}}}\pd{\tilde{\gamma_{km}}}{\gamma_{ij}},
} we obtain:
\labeq{ADMvsBSSN}{\pd{\ln\alpha}{\gamma_{ij}}=\Delta\frac{1}{12}\gamma^{ij}+
                               e^{-4\varphi}\Delta^{ij}-e^{-4\varphi}\Delta^{km}\frac{\gamma_{km}}{3}\gamma^{ij}.}
However, we are considering flat space so the latter becomes
\labeq{}{\pd{\ln\alpha}{\gamma_{ij}}=\Delta\frac{1}{12}\delta^{ij}+
                               \Delta^{ij}-\Delta^{km}\frac{\delta_{km}}{3}\delta^{ij}.}
Therefore, if we consider that the vector along which we perturb
is unit,
 the strong hyperbolicity condition of the ADM
formulation yields:
\labeq{}{{\pd{\ln\alpha}{\gamma_{ij}}v_iv_j=\frac{1}{12}(\Delta+
                               12\Delta^{ij}v_iv_j-4\Delta^{km}\delta_{km})}>0}
This last one is the same as condition \eqref{BSCondition2}.

If one considers gauges for which the lapse function depends on
the determinant of the three-metric and/or spacetime coordinates,
as we have already shown the condition for well-posedness, reduces
to
\labeq{}{A=\pd{\ln\alpha}{\ln\gamma}=\frac{1}{12}\Delta=\frac{1}{12}\pd{\ln\a}{\varphi}>0.
} which for both the ADM and BSSN language depends on the
functional form of the lapse only.

\section{Discussion and Conclusions}

We have presented a general theory to study the well-posedness of
constrained evolution of systems of quasi linear partial
differential equations, which consist of a set of $n$ first order
evolution equations and a set of $m$ first order constraint
equations with $m<n$. We applied this theory to constrained
evolution of 3+1 formulations of GR. In our analysis we explicitly
took into account the Hamiltonian and momentum constraints as well
as possible algebraic constraints on the evolution of
high-frequency perturbations of solutions of Einstein's equations.
Our analysis revealed  the existence of subsets of the linearized
Einstein's equations that control the well-posedness of
constrained evolution.

We demonstrated that the well-posedness of ADM and 3+1
formulations derived from ADM by adding combinations of
constraints to the right-hand-side (RHS) of ADM and/or by a linear
transformation of the dynamical ADM variables, depends entirely on
the properties of the gauge and are equivalent to ADM on the surface on constraints.
 We note that our method concerns the
constraint satisfying modes only. Those are present in free
evolution schemes, too. Therefore, a bad choice of gauge, which we
define as one that produces ill-posed constrained evolutions, is
bad for a free evolution, as well. However, a good choice of gauge
for a constrained evolution scheme cannot in principle guarantee
the well-posedness of a free evolution with the same gauge, due to
the existence of constraint violating modes.

Even on the surface of constraints we do not expect that all 3+1
formulation of GR which are derived from ADM by non-linear
transformations and addition of extra variables to have equivalent
well-posedness properties when using the same gauges. For example
the analysis of the exponential stretch rotation (ESR) formulation
\cite{24}, which is derived by a general non-linear exponential
transformation of the ADM variables, shows that in the simplest
case of geodesic slicing its behavior is elliptic, whereas ADM
with the same gauge is weakly hyperbolic. The analysis of ESR will
be the subject of a future paper.

In this paper we also analyzed the BSSN 3+1 formulation which is
derived by a non-linear transformation of the ADM variables and
addition of extra dynamical variables. We were able to show that
the well-posedness properties of BSSN and ADM on the surface of
constraints are similar for fixed and algebraic gauges. The
results seem to indicate that, in general,  the non-linear
transformation of variables leading from ADM to
 BSSN does not change
 the well-posedness properties of the constrained evolution when the same gauge is used.
However, the proof that the fully constrained evolution of BSSN is
well-posed if and only if the constrained evolution of ADM is
well-posed, if such proof exists, is out of the scope of this
paper.

Our study shows that fixed gauges, that is, when the lapse
function and the shift vector depend only on the spacetime
coordinates, result in an ill-posed Cauchy problem for the
constrained evolution of both ADM and BSSN as well as many other
3+1 formulations of GR. Algebraic gauges on the other hand can
give rise to a well-posed constrained evolution provided that they
satisfy \eqref{ADM-Condition-1}, \eqref{ADM-Condition-2} or
\eqref{BSCondition1}. In particular, fixed shift with the ``harmonic'' and
``$1+\log$'' slicing conditions, as well as with a densitized lapse
having $\sigma>0$ are all well behaved gauges. Our study of the
Bona-Masso family of hyperbolic slicing conditions showed that it
provides us with a well-posed constrained evolution. The study of
well-posedness of constrained evolution with maximal slicing and fixed shift shows
that it depends on the way this gauge is implemented. The
algebraic implementation $\gamma^{ij} K_{ij} = 0$ leads to a
well-posed evolution whereas the often used differential
implementation \eqref{Elliptic2} or \eqref{Elliptic} is ill-posed.
The parabolic extension of maximal slicing with fixed shift leads to an ill-posed
evolution. Finally, we demonstrated evidence that the constrained
evolution in conjunction with the "$\Gamma$-freezing" shift
condition and an algebraic lapse leads to a well behaved
constrained evolution at least in the case of 1D perturbations
about flat space. However, a complete well-posedness analysis is
still required and this will be a subject of a future paper.

Our analysis demonstrates that the weak hyperbolicity associated with
fixed gauges is directly related to the inertial deformation of a
synchronous reference frame with time which, in a general
non-linear case when the perturbations are not small, leads to the
formation of caustics (see equation \eqref{SubSet4b} and
discussion following it).

Finally, we note that gauge stability may be investigated more
directly by considering variations of gauge degrees of freedom
only. In general, this requires the analysis of a system of eight
quasi-linear partial differential equations presented in
\cite{11}.  The main advantage of the method outlined in this
paper is that, in addition to gauge conditions, it provides us
with subsets which control the constrained evolution of spacetime.
The method is also able to provide sufficient conditions of
well-posedness, whereas the analysis in \cite{11} gives only the
necessary conditions. The subsets controlling the constrained
evolution can be used for construction of stable numerical schemes
for 3+1 formulations of GR. This will be the subject of our future
paper.

\begin{acknowledgments}
V.P. thanks the Institute for Pure and Applied Mathematics at the
University of California Los Angeles for their hospitality and the
stimulating environment during his visit, A.K. thanks The
California Institute of Technology and in particular  Kip Thorne
and Joe Shepherd for hospitality during his sabbatical stay,  and
all authors thank Jakob Hansen for helpful discussions and
comments. Finally, we thank the anonymous referee for the
extremely useful comments that led to the improvement of the
manuscript.
\end{acknowledgments}

\appendix

\section{Strong Hyperbolicity for Constrained Evolution \label{appA}}

In this appendix we show that for uniformly diagonalizable systems
there exists a uniformly bounded similarity transformation $\hat
S$ which diagonalizes the principal matrix $\hat A$, provided that
the following two conditions are met: 1) $\hat A$ has real
eigenvalues, which are analytic functions of $v_k$, and 2) The
elements of $\hat A$ can be represented as ratios of analytic
functions of $v_k$.

To study well-posedness of a constrained evolution as outlined in
this paper we use the $m$ linearized constraint equations
\eqref{conlong1} to eliminate some of the dynamical variables.
Equations \eqref{conlong1} are an under-determined set of $m$
algebraic equations for the spatial derivatives of the $n$ unknown
variables, which in general can be solved for $m$ of the $n$
spatial derivatives of variables $\vec{u}$. We can write equations
\eqref{conlong1} schematically as follows
\labeq{conlong2}{\hat C(v_i)\pd{u_m}{\lambda}+\vec
F_m(v_i,\pd{u_{q}}{\lambda})=0,}
where $u_m$ is a column vector of $m$ of the $n$ dynamical
variables of the formulation, matrix $\hat C$ is $m\times m$ and
depends on the direction $v_i$ along which we perturb, and $\vec
F_m(v_i,\pd{u_{q}}{\lambda})$ is a column vector with $m$
components which are functions of the direction $v_i$ and the
spatial derivatives of the $q=n-m$ dynamical variables left, and
solve them as
\labeq{conlong3}{\pd{u_m}{\lambda}=-(\hat C)^{-1}\vec
F_m(v_i,\pd{u_{q}}{\lambda}).}
Substitution of \eqref{conlong3} in equations \eqref{deflong1}
leads to a set of $q=n-m$ linear partial differential equations
for $q$ of the initial $n$ variables, which is schematically given
by \eqref{minimal}. This elimination process includes inversion of
$\hat C(v_i)$, and thus division by its determinant, which is a
polynomial in the components of the unit one form $v_i$. This may
not be possible for every direction because there may be
directions which make matrix $\hat C$ singular. In order to obtain
the minimal set, the determinant $|\hat C|$ has to be
non-vanishing. The domain of a minimal set consists of all
directions for which $\vert \hat C \vert \neq 0$. For singular
directions $v_s$ we must use another set of $m$ dynamical
variables for which $|\hat C|\neq 0$ in the singular direction
$v_s$. It can be shown that this is always possible for the GR
equations.

A transformation matrix $S(v_i)$ which diagonalizes $\hat A$ in
\eqref{minimal} has the same domain as $\hat A$ and has non-zero
determinant in its domain. However, when we approach a singular
direction, the determinant of $S$ or its inverse may tend to zero
(or infinity) and then \eqref{detcon2} may not be satisfied.
However, the choice of eigenvectors and the corresponding
transformation matrix $\hat S$ are not unique. The eigenvectors
can be rescaled and this will change $\hat S$.

The systems analyzed in this paper for which the eigenvalues are
real and for which there exists a complete set of eigenvectors for
all directions, we find that the eigenvalues are analytic
functions of $v_k$ (see \eqref{gauge1-freq}, \eqref{gauge2-freq},
\eqref{bona-freq}). For such systems we show below that it is
always possible to rescale the eigenvectors in such a way that all
rescaled eigenvectors will be analytic functions of $v_k$. Then,
according to \cite{21} the transformation matrix $\hat S$ will
satisfy \eqref{detcon2} and thus the system will be strongly
hyperbolic and by definition well-posed.

First consider matrix $\hat A$. Its coefficients may be ratios of
polynomials due to substitution of constraints. This is the case
with the Einstein equations and all gauges we have studied in this
paper. We write this schematically as
\labeq{P1}{
                \left(\hat A(v_k)\right)_{ij} = \frac{p_{ij}(v_k)}{q_{ij}(v_k)} = \left( \begin{matrix}
                                                                                  \frac{p_{11}}{q_{11}} & ... & \frac{p_{1q}}{q_{1q}} \\
                                                                                  ...                   & ... & ... \\
                                                                                  \frac{p_{q1}}{q_{q1}} & ... & \frac{p_{qq}}{q_{qq}} \\
                                                                    \end{matrix}\right),
}
where $p_{ij}(v_k)$ and $q_{ij}(v_k)$ are polynomial (and hence
analytic) functions of $v_k$. We further assume that $\hat A_q$
has real eigenvalues and a complete set of eigenvectors for all
possible directions $v_k$ in its domain. This is the case for
algebraic gauges and the Bona-Masso hyperbolic gauges considered
in this paper.

Let $\vec{V}_l$ be a set of eigenvectors corresponding to the
eigenvalues $\omega_l$. They satisfy
\labeq{P2}{
                ( \hat A_q - \omega_l \hat I_q) \vec V_l \equiv \hat{\cal A}_l(v_k) \vec V_l = 0,
}
where $\hat I_q$ is the $q\times q$ identity matrix. The
coefficients of newly defined matrices $\hat {\cal A}_l$ are
ratios of analytic functions because  the eigenvalues of $\hat
A_q$ are analytic functions.
\labeq{P3}{
                  \left(\hat{ \cal A}(v_k)\right)_{ij} = \frac{P_{ij}(v_k)}{q_{ij}(v_k)} = \left( \begin{matrix}
                                                                                  \frac{P_{11}}{q_{11}} & ... & \frac{P_{1q}}{q_{1q}} \\
                                                                                  ...                   & ... & ... \\
                                                                                  \frac{P_{q1}}{q_{q1}} & ... & \frac{P_{qq}}{q_{qq}} \\
                                                                    \end{matrix}\right).
}
Consider now a particular eigenvalue, e.g.,  $\omega_1$ and a
particular eigenvector $\vec V_1$. Dropping the subscript~$1$ we
can write  \eqref{P2}  as
\labeq{P4}{
                     \begin{matrix}
                                     \frac{P_{11}}{q_{11}} V_1 + \frac{P_{12}}{q_{12}} V_2 + ... + \frac{P_{1q}}{q_{1q}} V_q = 0 \\
                                     ... \\
                                     \frac{P_{q1}}{q_{q1}} V_1 + \frac{P_{q2}}{q_{q2}} V_2 + ... + \frac{P_{qq}}{q_{qq}} V_q = 0 \\
                     \end{matrix},
}
here subscripts of $V$ indicate components of a particular
eigenvector $\vec V= \vec V_1$ we consider. If the rank of  the
matrix $\hat {\cal A}_l$ is $r$, where $r<q$ then there are only
$r$ linearly independent equations in \eqref{P4}. This means that
we can solve the algebraic set for $r$ of the $q$ components of
the eigenvector $\vec V$, which we denote by $\vec V_r$. The
remaining $s=q-r$ components of $\vec V$ form a vector $\vec V_s$.
We write the reduced system schematically as
\labeq{P8}{\hat D_r \vec V_r= \hat D_s \vec V_s \equiv \vec B_r, }
where $\hat D_r$ is a $r\times r$ matrix and $\hat D_s$ is a
$r\times s$ matrix, which are both formed by elements of ${\cal
A}_l$. Therefore, both $\hat D_r$ and $\hat D_s$ have elements
which are ratios of analytic functions of $v_k$. The non-zero
components of $\vec V_s$ can be chosen freely, for example as
constants, so that components of $\vec B_r$ will be ratios of
analytic functions in $v_k$. By Cramer's rule the solution of
system \eqref{P8} for the $r$ unknown eigenvector components is
\labeq{P9}{(\vec V_r)_i=\frac{|(\hat D_r)_i|}{|\hat D_r|}, } where
$(\vec V_r)_i$ is the $i$-th component of $\vec V_r$  and $(\hat
D_r)_i$ is the matrix formed by replacing the $i$-th column of
$\hat D_r$ by the column vector $\vec B_r$. Therefore, all
components of the  eigenvector $\vec V$ can be expressed as ratios
of analytic functions, since the sums and products of analytic
functions are analytic, which we write schematically as
\labeq{P10}{(\vec V)_i=\frac{P_i(v_k)}{Q_i(v_k)}. }
If we multiply the  eigenvector $ \vec V$ by  the product $Q =
\Pi_i Q_i(v_k)$ the rescaled  eigenvector will be an analytic
function of $v_k$. We can carry out such rescaling for all
eigenvectors. Then according to \cite{21} \eqref{conlong2} is
strongly hyperbolic because $\hat A$ has real eigenvalues and a
complete set of eigenvectors which are analytic functions of $v_k$
for any direction.

We now illustrate an example of rescaling that leads to a
uniformly bounded $\hat S$. Consider perturbations in the
$x^1,x^2$ plane about flat spacetime for the ADM + densitized
lapse system ($\alpha=C(x_i)\gamma^\sigma$), for which the
components $v_1 \mbox{ and } v_2$, of the unit one-form along
which we perturb, are assumed non-vanishing. For this simple case
q=30-22=8. The linearized evolution equations of a minimal set are
in matrix notation:
\labeq{}{  \pd{\vec{u}}{t} = A_8 \pd{\vec{u}}{\lambda},} where
\labeq{}{  \vec{u}^{T}=(\begin{array}{ccccccccc}
K_{11} & K_{23} & K_{33} & D_{11;1} & D_{13;1} & D_{22;1} &
D_{23;1} & D_{33;1}
\end{array})
}
and
\begin{widetext}
\labeq{}{A_8= \left(
\begin{array}{ccccccccc}
0 & 0 & 0 & -\alpha v_1 \sigma  & 0 & -\alpha v_1 \sigma  & 0 &
\frac{\alpha \left[1-v_1{}^2 (2 \sigma +1)\right]}{2 v_1}   \\
0 & 0 & 0 & 0 & \frac{\alpha v_2}{2} & 0 & -\frac{\alpha v_1}{2} & 0 \\
0 & 0 & 0 & 0 & 0 & 0 & 0 & -\frac{\alpha}{2 v_1} \\
-2 \alpha v_1 & 0 & 0 & 0 & 0 & 0 & 0 & 0 \\
0 & 2 \alpha v_2 & 0 & 0 & 0 & 0 & 0 & 0 \\
-\frac{2 \alpha v_2{}^2}{v_1} & 0 & -\frac{2 \alpha \left(1-2 v_1^2\right)}{v_1} & 0 & 0 & 0 & 0 & 0 \\
0 & -2 \alpha v_1 & 0 & 0 & 0 & 0 & 0 & 0 \\
0 & 0 & -2 \alpha v_1 & 0 & 0 & 0 & 0 & 0
\end{array}
\right), }\\
\end{widetext}
where $\vec{u}^{T}$ is the transpose of $\vec{u}$. When
$\sigma>0$, this matrix always has real eigenvalues
$\left\{0,0,-\alpha,\alpha,-\alpha,\alpha, -\alpha \sqrt{2
\sigma},\alpha \sqrt{2 \sigma}\right\}$
  and complete set of
eigenvectors in its domain, i.e. $v_1\neq 0,\ v_2\neq 0$. This is
also the domain of the matrix ($S$) of eigenvectors of $A_8$ as
columns, which is a matrix that diagonalizes $A_8$ via a
similarity transformation. The determinant of this matrix turns
out to be:
\labeq{}{|S|= -\frac{\sqrt{2 \sigma }}{v_1{}^2 v_2{}^5 }.}
As we approach the singular points $v_1=0,\ v_2=0$ this
determinant blows up and one cannot obtain an upper bound to
satisfy \eqref{detcon2}. However, if we define a new matrix
(within the same domain) by
\labeq{}{\tilde{S}=v_1{}^{2/8}v_2{}^{5/8} S,}
the determinant of this new matrix is
\labeq{}{|\tilde{S}|= -\sqrt{2 \sigma }.}
This is a well behaved non-singular transformation  that satisfies
\eqref{detcon2}.


\section{Ill-posedness of Parabolic Maximal Slicing \label{appB}}
\vspace{-0.3cm} In this appendix we demonstrate that the ADM
constrained evolution with parabolic maximal slicing is ill-posed.
Since the parabolic maximal slicing is a second order differential
equation we introduce new variables to achieve first order form at
least to the evolution equations of those variables, whose
solution may determine the solution of the entire system. Let
$A_k=\pd{\a}{x^k}$. The evolution equations for these variables
can be easily obtained by commuting a time with a space derivative
and after we perturb along a specified direction about a base
solution and make use of the definitions of these variables, the
equations that describe the evolution of high frequency
perturbations of $A_k$ are (assuming without any loss of
generality that the component of ${\mathbf v}$ along $x^1$ is
non-zero):
\begin{widetext}
\labeq{A_k}{\begin{split}\pd{\delta A_k}{t} =&
\frac{1}{\epsilon}\bigg[\frac{1}{v_1}v_k\pdd{\delta A_1}{\lambda}
-\gamma^{ij}\Gamma^{\ell}{}_{ij}\frac{v_k v_\ell}{v_1}\pd{\delta
A_1}{\lambda}- \half v_k\gamma^{ij}\gamma^{\ell n}
A_\ell\left(2\pd{D_{in;j}}{\lambda}-\pd{D_{ij;n}}{\lambda}\right)\\
& \hspace{4cm}-\alpha \g^{im}\g^{j\ell}K_{ij}v_k\pd{\delta K_{\ell
m}}{\l}-\alpha K^{ij}v_k \pd{\delta K_{ij}}{\lambda}-c\g^{ij}v_k
\pd{\delta K_{ij}}{\lambda}\bigg].
\end{split}}
\end{widetext}
Now define ${\cal K}_{ij}\equiv K_{ij}+\epsilon v_1 v_i v_j
\alpha$. Then equations \eqref{Parabolic}-\eqref{dDijkParabolic}
yield for the evolution of high frequency perturbations of $\a,
{\cal K}_{ij} \mbox{ and } D_{ij;k}$:

\labeq{alpha}{\pd{\delta\a}{t}=\frac{1}{\epsilon v_1}\pd{\delta
A_1}{\l},}
\labeq{ADMK}{\begin{split} \pd{\delta {\cal K}_{ij}}{t}=& \a R^1_{ij}, \\
\pd{\delta D_{ij;k}}{t}= & -2\a v_k \pd{\delta {\cal K}_{ij}}{\l}.
\end{split}}

The momentum constraints \eqref{MD} have exactly the same form for
this newly defined ``extrinsic curvature" variable ${\cal K}_{ij}$
in the limit of high frequency  perturbations, i.e.
\labeq{MCmod}{\quad   \gamma^{ms}v_s  \pd{\delta {\cal
K}_{mi}}{\lambda} -
          v_i \gamma^{mn} \pd{\delta {\cal K}_{mn}}{\lambda}  = 0.}
System \eqref{A_k}-\eqref{ADMK} has now decoupled in two parts.
One consists of the evolution equations for the $A_k$ variables
and the lapse function, equations \eqref{A_k} and \eqref{alpha}.
The other part consists of the evolution equations for $D_{ij;k}$
and ${\cal K}_{ij}$, equations \eqref{ADMK}, governed by the exact
same constraint equations as variables $D_{ij;k}$ and $K_{ij}$.
The constrained evolution of this last set has been analyzed in
the section of fixed gauges and ADM and has been found to be
ill-posed. Since the constrained solution of \eqref{ADMK} defines
the solution of \eqref{A_k} and \eqref{alpha}, the entire system
is ill-posed. And hence the constrained evolution with the
parabolic extension of maximal slicing  is weakly hyperbolic and
hence is ill-posed.

\end{document}